\documentclass[12pt,english]{article}
\usepackage[latin9]{inputenc}
\usepackage[a4paper]{geometry}
\usepackage{textcomp}
\usepackage{amsmath}
\usepackage{graphicx}
\usepackage[authoryear]{natbib}
\usepackage{subscript}

\makeatletter

\DeclareTextSymbolDefault{\textquotedbl}{T1}
\providecommand{\tabularnewline}{\\}

\newcommand{\lyxaddress}[1]{
	\par {\raggedright #1
	\vspace{1.4em}
	\noindent\par}
}

\@ifundefined{date}{}{\date{}}
\makeatother

\usepackage{babel}
\begin{document}
\title{Economics of carbon-dioxide abatement under an exogenous constraint
on cumulative emissions }
\author{Ashwin K Seshadri}
\maketitle

\lyxaddress{Centre for Atmospheric and Oceanic Sciences and Divecha Centre for
Climate Change, Indian Institute of Science, Bangalore, 560012, India
(ashwin@fastmail.fm; ashwins@iisc.ac.in)}
\begin{abstract}
The fossil-fuel induced contribution to further warming over the 21st
century will be determined largely by integrated CO\textsubscript{2}
emissions over time rather than the precise timing of the emissions,
with a relation of near-proportionality between global warming and
cumulative CO\textsubscript{2} emissions. This paper examines optimal
abatement pathways under an exogenous constraint on cumulative emissions.
Least cost abatement pathways have carbon tax rising at the risk-free
interest rate, but if endogenous learning or climate damage costs
are included in the analysis, the carbon tax grows more slowly. The
inclusion of damage costs in the optimization leads to a higher initial
carbon tax, whereas the effect of learning depends on whether it appears
as an additive or multiplicative contribution to the marginal cost
curve. Multiplicative models are common in the literature and lead
to delayed abatement and a smaller initial tax. The required initial
carbon tax increases with the cumulative abatement goal and is higher
for lower interest rates. Delaying the start of abatement is costly
owing to the increasing marginal abatement cost. Lower interest rates
lead to higher relative costs of delaying abatement because these
induce higher abatement rates early on. The fraction of business-as-usual
emissions (BAU) avoided in optimal pathways increases for low interest
rates and rapid growth of the abatement cost curve, which allows a
lower threshold global warming goal to become attainable without overshoot
in temperature. Each year of delay in starting abatement raises this
threshold by an increasing amount, because the abatement rate increases
exponentially with time. 
\end{abstract}

\section{Introduction}

Abatement of global warming requires reductions in emissions of carbon-dioxide
(CO\textsubscript{2}), with estimated costs of large-scale decarbonization
varying substantially across studies and models (\citet{Manne1993,Grubb1993,Weyant1993}),
a major factor being baseline energy demand under \textquotedblleft business
as usual\textquotedblright{} (BAU) (\citet{Rogelj2015}). An important
context of abatement is approximate proportionality between global
warming from CO\textsubscript{2} and its cumulative emissions (\citet{Allen2009,Matthews2009}),
and more generally that this relation is independent of emissions
pathway (\citet{Herrington2014,Seshadri2017a}). Such \textquotedblleft path-independence\textquotedblright{}
simplifies comparison of CO\textsubscript{2} emissions scenarios,
requiring only that cumulative emissions be considered (\citet{Zickfeld2009,Bowerman2011}).
In general, since global warming is not in equilibrium with radiative
forcing owing to the slow climate response (\citet{Stouffer2004,Held2010}),
comparison of emissions scenarios invokes climate modeling uncertainties
(\citet{Shine2005,Allen2016,Seshadri2017}) and this is where cumulative
emissions accounting presents a major simplification for CO\textsubscript{2}
(\citet{Matthews2012,Stocker2013a}). In this case, path independence
arises from the direct effects of changing cumulative emissions being
an order of magnitude larger than effects of the slowly changing airborne
fraction of excess CO\textsubscript{2}, which can be understood in
terms of corresponding timescales of evolution (\citet{Seshadri2017a}).

Cumulative carbon budgets also raise new questions for analyses of
mitigation pathways, given an exogenous constraint on cumulative emissions,
in contrast with the effects of a Pigouvian tax (\citet{Pigou1920}),
i.e. a tax that internalizes the external damage costs of emissions.
The evolution of such a tax must evolve according to the changing
social cost of carbon (SCC) (\citet{Pearce2003,Nordhaus2017}), i.e.
the discounted future global damages from the pulse emission of a
ton of CO\textsubscript{2} . Since such a tax is meant to internalize
the present value of the future costs of present emissions, the value
of estimated SCC would determine the level of abatement. In contrast
the scientific discussion about cumulative carbon budgets has examined
decarbonization strategies achieving stringent global warming goals
(e.g. \citet{Rogelj2011,Millar2017}). These do not directly reference
the SCC, but are motivated by a precautionary approach to avoiding
large but uncertain human and environmental consequences of failure
(\citet{Meinshausen2009,Hoegh-Guldberg2018}). Moreover, although
the effectiveness of the earlier Kyoto protocol is marred by limited
scope (\citet{Shishlov2016}), growing evidence of the temperature-sensitivity
of climate change impacts has animated rising ambition (\citet{Smith2009,Hoegh-Guldberg2018}),
translating to difficult cumulative emissions goals (\citet{Meinshausen2009,Stocker2013,Rhein2013}),
now set in the context of the Paris Agreement seeking to hold \textquotedblleft increase
in the global average temperature to well below 2�C above pre-industrial
levels and pursuing efforts to limit the temperature increase to 1.5�C
above pre-industrial levels, recognizing that this would significantly
reduce the risks and impacts of climate change\textquotedblright{}
(\citet{UNFCCC2015}). 

Previous studies have considered economics of cumulative emissions
abatement, including optimal paths that optimize a social welfare
functional, as well as abatement expenditures under an exogenous constraint
on cumulative emissions (\citet{Dietz2019,Emmerling2019,Ploeg2019}).
These studies have undertaken far-reaching analyses of wide-ranging
questions such as optimal temperatures (\citet{Dietz2019}), carbon
price pathways (\citet{Dietz2019,Emmerling2019,Ploeg2019}), effect
of damage costs (\citet{Dietz2019}), effects of discounting (\citet{Emmerling2019}),
carbon budget overshoots (\citet{Emmerling2019}), inter-generational
inequality aversion (\citet{Ploeg2019}), and exogenous technical
progress (\citet{Ploeg2019}). The present paper focuses on abatement
under an exogenous cumulative emissions constraint, extending the
previous analyses to model BAU emissions, the effects of endogenous
learning in addition to climate damages, and allowing economic growth
and interest rates to vary with time. Prior authors have examined
endogenous learning, starting with \citet{Arrow1962}, who modeled
it as depending on cumulative investment and, although not a standard
part of IAMs, endogenous learning has been studied in the CO\textsubscript{2}
abatement context (e.g. \citet{Goulder2000,Zwaan2002,Wing2006,Gillingham2008}
), although effects on optimizing pathways under a cumulative emissions
constraint have not been examined. We also examine the effects of
a delayed start to abatement. Cumulative emissions budgets might appear
ambivalent for stringent climate policy, with delayed present abatement
manageable via increases in future abatement. We examine the costs
of delaying the onset of a global carbon tax and we show that the
tradeoff between lower discounted future costs versus the need for
a rapid transition resulting from a delay is limited once the time-value
of money is factored into the optimal abatement pathways. The need
for more rapid abatement following a delay drives its cost. 

The present work examines cost abatement pathways and consequences
for the carbon tax, assuming an exogenously determined end-of-century
cumulative emissions goal. These pathways minimize abatement costs,
either with or without endogenous learning, and we also consider the
effects of including climate damages in the optimization. Where we
address similar questions, the framework is similar to the aforementioned
studies (\citet{Dietz2019,Emmerling2019}), wherein minimization gives
rise to a two-point boundary value problem on cumulative abatement,
which can solved for evolution of abatement rates. The corresponding
evolution of a carbon tax in the context of a cumulative emissions
constraint is examined, and the results correspond to those of the
previous studies (\citet{Dietz2019,Emmerling2019}) for cases where
endogenous learning\textquoteright s effects are omitted. We consider
applications to related questions: the influences on the initial value
of the carbon tax, the effect of delaying abatement on growth in the
present value of total expenditures as well as growth of the initial
carbon tax, and the economics of avoiding overshoot in the global
warming trajectory. Overshoot scenarios of the carbon budget have
been considered previously by \citet{Emmerling2019}. While the exogenous
specification of cumulative emissions does not prevent overshoot of
global warming compared to its value at the end of the time-horizon,
limiting the magnitude of overshoot is important for averting impacts
triggered by temperature thresholds (\citet{Smith2009,Hoegh-Guldberg2018}).
Therefore we examine the constraints on the minimum amount of global
warming that can be achieved without any overshoot and, given its
approximate proportionality with cumulative emissions, avoiding overshoot
of global warming requires net emissions to not be negative. 

\section{Models and methods}

We are given an exogenous cumulative abatement goal $M_{tot}$, the
total reduction in emissions compared to the business-as-usual scenario.
Adopting a continuous-time formulation, with $t=0$ denoting the present,
and $m\left(t\right)$ being the abatement rate in year $t$, the
cumulative abatement $M\left(t\right)=\int_{0}^{t}m\left(s\right)ds$
must satisfy $M\left(T\right)=M_{tot}$ at the end of the time-horizon
$t=T$. Following the notation typically used in variational calculus,
abatement rate is henceforth denoted as $\dot{M}\left(t\right)\equiv\frac{dM\left(t\right)}{dt}$
instead of the above $m\left(t\right)$. The average cost of emissions
abatement, in the absence of endogenous learning, is an increasing
function $\gamma_{0}\left(\dot{M}\left(t\right)\right)$ of abatement
rate, assuming that each year the cheaper abatements are undertaken
first. Previous authors have shown evidence in a number of sectors
for cost reductions depending on cumulative production (``endogenous
learning'' or ``learning by doing'') (\citet{Rubin2004,Nagy2013}).With
endogenous learning, the average cost is lowered progressively as
cumulative abatement $M\left(t\right)$ grows, so that in general
the average cost is $\gamma\left(\dot{M}\left(t\right),M\left(t\right)\right)$.
Climate damages are modeled as an increasing function of global warming. 

\subsection{Business-as-usual emissions}

The business as usual (BAU) emissions rate depends on global GDP $G\left(t\right)$
and global emissions intensity $\mu\left(t\right)$ of CO\textsubscript{2}
as $\dot{E}_{BAU}\left(t\right)=G\left(t\right)\mu\left(t\right)$.
Its growth rate depends on the global income elasticity of emissions
$\theta$ since, over time-interval $\triangle t$ , the relative
change in BAU emissions rate is $\triangle\dot{E}_{BAU}/\dot{E}_{BAU}=\theta\triangle G/G$.
Thus, assuming exogenous GDP growth rate $r\left(t\right)=\triangle G\left(t\right)/G\left(t\right)$,\footnote{Absence of mitigation can also adversely impact economic growth. Quoting
\citet{Stern2016}, \textquotedbl So the business-as-usual baseline,
against which costs of action are measured, conveys a profoundly misleading
message to policymakers that there is an alternative option in which
fossil fuels are consumed in ever greater quantities without any negative
consequences to growth itself.\textquotedbl{} Treating this mathematically
is beyond our scope. }emissions intensity under BAU declines at the rate $-\frac{1}{\mu}\frac{d\mu}{dt}=\left(1-\theta\right)\frac{r}{1+r}$
(Appendix 1). Since we can approximate $1/\left(1+r\right)\cong1$,
the emissions intensity under BAU can be described by $\mu\left(t\right)=\mu\left(0\right)e^{\left(\theta-1\right)R\left(t\right)}$
where $R\left(t\right)=\int_{0}^{t}r\left(s\right)ds$ is the integrated
GDP growth rate and with the GDP growing as $G\left(t\right)=G\left(0\right)e^{R\left(t\right)},$the
annual emissions rate under BAU is
\begin{equation}
\dot{E}_{BAU}\left(t\right)=\dot{E}_{BAU}\left(0\right)e^{\theta R\left(t\right)}\label{eq:p1}
\end{equation}
where we have allowed for time-varying $r$ but assume constant $\theta$.
Across nations, the income elasticity of emissions varies considerably,
depending on the pattern of economic growth, being generally smaller
for richer economies (\citet{Webster2008}) and thus liable to decrease
as economies develop. 

\subsection{Abatement cost model in the absence of endogenous learning}

The average cost of emissions abatement is an increasing function
of the abatement rate 
\begin{equation}
\gamma_{0}\left(\dot{M}\left(t\right)\right)=c_{0}+c_{1}\left(\frac{\dot{M}\left(t\right)}{\dot{M}_{max}\left(t\right)}\right)^{c_{2}}\label{eq:p2}
\end{equation}
where $c_{1}>0$ and $c_{2}>0$, but the intercept $c_{0}$ can be
negative to reflect the availability of negative-cost techniques.
$\dot{M}_{max}\left(t\right)$ describes the horizontal scale of the
abatement graph, at which average cost is fixed at $c_{0}+c_{1}$,
which we take as equal to the BAU emissions $\dot{M}_{max}\left(t\right)=\dot{E}_{BAU}\left(t\right)$.
As BAU emissions grow, abatement possibilities are assumed to expand
proportionally. This also leads to the result that the abatement cost
remains constant in the absence of policy-induced decarbonization
or exogenous cost reductions, in line with models such as DICE (\citet{Nordhaus2013}).
Higher emissions reductions than $\dot{E}_{BAU}\left(t\right)$ are
possible, being in the realm of net-negative emissions, with resulting
average costs following Eq. (\ref{eq:p2}) exceeding $c_{0}+c_{1}$.
The model parallels prior approaches based on the marginal abatement
cost, which is taken to be some power of the emissions reductions
relative to BAU, e.g. \citet{Emmerling2019}. It assumes that abatement
possibilities are independent of the past history of mitigation, depending
only on the BAU pathway of emissions.

\subsection{Endogenous learning model}

Only learning by doing is modeled, not exogenous learning with time,
influencing the average cost as
\begin{equation}
\gamma\left(\dot{M}\left(t\right),M\left(t\right)\right)=\left(\gamma_{0}\left(\dot{M}\left(t\right)\right)-f\left(M\left(t\right)\right)\right)h\left(M\left(t\right)\right)\label{eq:p3}
\end{equation}
where $\gamma_{0}\left(\dot{M}\left(t\right)\right)$ describes the
cost curve in the absence of learning, and the terms $f\left(M\left(t\right)\right)$
and $h\left(M\left(t\right)\right)$ are introduced to admit the different
possibilities: learning shifts the cost curve down, scales it down,
or does both. If there is only an additive term, then $f\left(M\left(t\right)\right)\geq0$
and is increasing in $M\left(t\right)$, while $h\left(M\left(t\right)\right)=1$
and is fixed. If there is only a multiplicative term, then $f\left(M\left(t\right)\right)=0$
and is fixed while $h\left(M\left(t\right)\right)$, satisfying $0<h\left(M\left(t\right)\right)\leq1$,
is a decreasing function of cumulative abatement. The functions must
satisfy $f\left(0\right)=0$ and $h\left(0\right)=1$. For example,
an additive effect might take form $f\left(M\left(t\right)\right)=c_{f}M\left(t\right)$
with $c_{f}>0$ and $h\left(M\left(t\right)\right)=1$. A multiplicative
effect could be $h\left(M\left(t\right)\right)=e^{-M\left(t\right)/M_{h}}$
with constant parameter $M_{h}>0$. In contrast to this decreasing
exponential model, a power-law model suggested by the literature (\citet{Rubin2004,Nagy2013})
has form $h\left(M\left(t\right)\right)=\left(M\left(t\right)/M\left(t=0\right)\right)^{-b}$
with necessarily $M\left(t=0\right)>0$ . Any combination of the above
effects is also possible. 

\subsection{Climate damages model}

Climate damages are modeled as a fraction of GDP
\begin{equation}
D\left(t\right)=d\left(t\right)G\left(t\right)\label{eq:p4}
\end{equation}
where damage fraction $d\left(t\right)$ is increasing in global warming
$\triangle T\left(t\right)$, which in turn is proportional to cumulative
emissions
\begin{equation}
\triangle T\left(t\right)=\alpha E\left(t\right)\label{eq:p5}
\end{equation}
with $\alpha$ being the ``transient climate response to cumulative
carbon emissions'' (TCRE) (\citet{Matthews2018}). Treating the damage
fraction as a function of the temperature at time $t$ is conventional,
but neglects important factors such as delayed impacts of sea-level
rise and lasting impacts of extreme events. Cumulative emissions is
measured from preindustrial time $-t_{PI}$ 
\begin{equation}
E\left(t\right)=\int_{-t_{PI}}^{0}\dot{E}\left(s\right)ds+\int_{0}^{t}\dot{E}\left(s\right)ds\label{eq:p6}
\end{equation}
and, since the emissions rate is
\begin{equation}
\dot{E}\left(t\right)=\dot{E}_{BAU}\left(t\right)-\dot{M}\left(t\right)\label{eq:p7}
\end{equation}
where $\dot{E}_{BAU}\left(t\right)$ is the rate under ``business
as usual'' (BAU), we obtain for cumulative emissions 
\begin{equation}
E\left(t\right)=E_{BAU}\left(t\right)-M\left(t\right)\label{eq:p8}
\end{equation}
in terms of $E_{BAU}\left(t\right)$, the cumulative emissions under
BAU. With respect to cumulative abatement, global warming is $\triangle T\left(t\right)=\alpha\left\{ E_{BAU}\left(t\right)-M\left(t\right)\right\} $
and we subsequently treat alternate formulations of damage function
$d\left(\triangle T\right).$

\subsection{Constrained optimization framework}

The analysis of this paper is founded on pathways minimizing the present
value of abatement costs, or sum of total abatement and damage costs.
Since $\gamma\left(t\right)$ is average cost per unit of abatement
in year $t$, total abatement cost in a year is
\begin{equation}
A\left(M\left(t\right),\dot{M}\left(t\right)\right)=\gamma\left(t\right)\dot{M}\left(t\right)=\left(\gamma_{0}\left(\dot{M}\left(t\right)\right)-f\left(M\left(t\right)\right)\right)h\left(M\left(t\right)\right)\dot{M}\left(t\right)\label{eq:p9}
\end{equation}
Similarly, the damage cost is given by Eq. (\ref{eq:p4}). We must
minimize the present value of these costs, between $t=0$ and $t=T$,
subject to the aforementioned constraint on cumulative abatement $\int_{0}^{T}\dot{M}\left(s\right)ds=M_{tot}.$
At risk-free interest rate $i\left(t\right)$, the present value of
the cost is
\begin{equation}
C_{total}=\int_{0}^{T}e^{-I\left(t\right)}C\left(t,M\left(t\right),\dot{M}\left(t\right)\right)dt\label{eq:p10}
\end{equation}
where 
\begin{multline}
C\left(t,M\left(t\right),\dot{M}\left(t\right)\right)=\left(c_{0}+c_{1}\left(\frac{\dot{M}\left(t\right)}{\dot{M}_{max}\left(t\right)}\right)^{c_{2}}-f\left(M\left(t\right)\right)\right)h\left(M\left(t\right)\right)\dot{M}\left(t\right)\\
+d\left(M\left(t\right)\right)G\left(t\right)\label{eq:p11}
\end{multline}
is the sum of abatement and damage costs and $I\left(t\right)=\int_{0}^{t}i\left(s\right)ds$
is the cumulative interest rate at time $t$. Minimizing the functional
in Eq. (\ref{eq:p10}) subject to constraint on cumulative abatement,
for fixed time-horizon $T$, involves finding a solution to the Euler-Lagrange
equation (Appendix 2)

\begin{equation}
\frac{\partial g}{\partial M}+\lambda\frac{\partial\dot{M}}{\partial M}=\frac{d}{dt}\left(\frac{\partial g}{\partial\dot{M}}+\lambda\frac{\partial\dot{M}}{\partial\dot{M}}\right)\label{eq:p12}
\end{equation}
where 
\begin{equation}
g\left(t,M\left(t\right),\dot{M}\left(t\right)\right)=e^{-I\left(t\right)}C\left(t,M\left(t\right),\dot{M}\left(t\right)\right)\label{eq:p13}
\end{equation}
is the integrand in Eq. (\ref{eq:p10}). Tables 1 and 2 list some
of the frequently used emissions and climate parameters, and economic
parameters, respectively. 

\pagebreak{}

Table 1: Emissions and climate parameters

\begin{tabular}{|c|c|c|}
\hline 
Symbol & Description & Units\tabularnewline
\hline 
\hline 
$E\left(t\right)$ & cumulative emissions of CO\textsubscript{2} & Gton\tabularnewline
\hline 
$M\left(t\right)$ & cumulative abatement of CO\textsubscript{2} & Gton\tabularnewline
\hline 
$\dot{M}\left(t\right)$ & abatement rate & Gton year\textsuperscript{$-1$}\tabularnewline
\hline 
$\dot{M}_{max}\left(t\right)$ & maximum abatement rate, equal to BAU emissions & Gton year\textsuperscript{$-1$}\tabularnewline
\hline 
$r_{BAU}\left(t\right)$ & growth rate of BAU emissions & year\textsuperscript{$-1$}\tabularnewline
\hline 
$\triangle T\left(t\right)$ & global warming from CO\textsubscript{2} & K\tabularnewline
\hline 
$\alpha$ & transient climate response to cumulative carbon emissions (TCRE) & K Gton\textsuperscript{$-1$}\tabularnewline
\hline 
$M_{tot}$ & cumulative abatement & Gton\tabularnewline
\hline 
\end{tabular}

\pagebreak{}

Table 2: Economics parameters

\begin{tabular}{|c|c|c|}
\hline 
Symbol & Description & Units\tabularnewline
\hline 
\hline 
$P$ & carbon tax & $\$$ ton\textsuperscript{$-1$}\tabularnewline
\hline 
$i\left(t\right)$ & discount rate & year\textsuperscript{$-1$}\tabularnewline
\hline 
$I\left(t\right)$ & cumulative interest rate & dimensionless\tabularnewline
\hline 
$\theta$ & income elasticity of emissions & dimensionless\tabularnewline
\hline 
$\gamma_{0}\left(\dot{M}\left(t\right),M\left(t\right)\right)$ & average cost of abatement without learning & trillion \$ Gton\textsuperscript{$-1$} \tabularnewline
\hline 
$\gamma\left(\dot{M}\left(t\right),M\left(t\right)\right)$ & average cost of abatement with learning & trillion \$ Gton\textsuperscript{$-1$} \tabularnewline
\hline 
$\beta\left(\dot{M}\left(t\right),M\left(t\right)\right)$ & marginal cost of abatement with learning & trillion \$ Gton\textsuperscript{$-1$} \tabularnewline
\hline 
$A\left(t\right)$ & abatement cost in year $t$ & trillion \$\tabularnewline
\hline 
$c_{0}$ & intercept of average cost function & trillion \$ Gton\textsuperscript{$-1$} \tabularnewline
\hline 
$c_{1}$ & coefficient of average cost function & trillion \$ Gton\textsuperscript{$-1$} \tabularnewline
\hline 
$c_{2}$ & exponent of average cost function & dimensionless\tabularnewline
\hline 
$G\left(t\right)$ & global GDP & trillion \$\tabularnewline
\hline 
$d\left(t\right)$ & climate damage fraction & dimensionless\tabularnewline
\hline 
$D\left(t\right)$ & climate damage cost & trillion \$\tabularnewline
\hline 
$T_{0}$ & reference global warming for power-law damage-fraction & $10$ K\tabularnewline
\hline 
$d_{0}$ & power-law damage fraction at $\triangle T=T_{0}$ & dimensionless\tabularnewline
\hline 
$d_{1}$ & exponent of power-law damage function & dimensionless\tabularnewline
\hline 
$d_{2}$ & coefficient of logistic damage function & dimensionless\tabularnewline
\hline 
$E_{D}$ & parameter for logistic damage function  & Gton\tabularnewline
\hline 
$f\left(M\left(t\right)\right)$ & additive learning & trillion \$ Gton\textsuperscript{$-1$}\tabularnewline
\hline 
$h\left(M\left(t\right)\right)$ & multiplicative learning & dimensionless\tabularnewline
\hline 
$M_{h}$ & parameter for exponential model of learning & Gton\tabularnewline
\hline 
$-b$ & exponent of power-law model of learning & dimensionless\tabularnewline
\hline 
$c_{f}$ & coefficient for additive model of learning & trillion \$ Gton\textsuperscript{$-2$}\tabularnewline
\hline 
\end{tabular}

\pagebreak{}

\section{Results}

\subsection{Optimal abatement rate}

Using Eqs. (\ref{eq:p11}) and (\ref{eq:p13}), and simplifying (Appendix
3), the Euler-Lagrange Eq. (\ref{eq:p12}) for the evolution of the
abatement rate $\dot{M}\left(t\right)$ becomes
\begin{multline}
c_{1}c_{2}\left(c_{2}+1\right)\left(\frac{\dot{M}\left(t\right)}{\dot{M}_{max}\left(t\right)}\right)^{c_{2}}\frac{\ddot{M}\left(t\right)}{\dot{M}\left(t\right)}=i\left(t\right)\left\{ c_{0}+c_{1}\left(c_{2}+1\right)\left(\frac{\dot{M}\left(t\right)}{\dot{M}_{max}\left(t\right)}\right)^{c_{2}}-f\left(M\left(t\right)\right)\right\} \\
+c_{1}c_{2}\left(c_{2}+1\right)\left(\frac{\dot{M}\left(t\right)}{\dot{M}_{max}\left(t\right)}\right)^{c_{2}}r_{BAU}\left(t\right)-\frac{c_{1}c_{2}}{\dot{M}_{max}^{c_{2}}\left(t\right)}\dot{M}\left(t\right)^{c_{2}+1}\frac{h'\left(M\left(t\right)\right)}{h\left(M\left(t\right)\right)}+d'\left(M\left(t\right)\right)\frac{G\left(t\right)}{h\left(M\left(t\right)\right)}\label{eq:p14}
\end{multline}
where $r_{BAU}\left(t\right)=\frac{1}{\dot{M}_{max}\left(t\right)}\left\{ \frac{d}{dt}\dot{M}_{max}\left(t\right)\right\} =\theta r\left(t\right)$
is the growth rate of BAU emissions. This 2\textsuperscript{nd}-order
equation in $M\left(t\right)$ requires two boundary values. At the
present time when abatement is to begin, we have $M\left(t=0\right)=0$,
\footnote{Except when considering the power-law multiplicative model of endogenous
learning, which requires a small value of $M\left(0\right)>0$. } in addition to cumulative abatement $M\left(t=T\right)=M_{tot}$.
The parameters appearing in Eq. (\ref{eq:p14}), along with these
boundary conditions, uniquely prescribe the abatement pathway. We
assume henceforth that average costs are increasing, i.e. $c_{1}$
and $c_{2}$ are positive; for, if average cost function were constant,
with either $c_{1}$ or $c_{2}$ being equal to zero, we obtain an
algebraic equation in $M\left(t\right)$, which in general is not
soluble. \footnote{If damages are not counted it becomes $i\left(t\right)\left(c_{0}-f\left(M\left(t\right)\right)\right)=0$.
This equation, for $i\neq0$ yields $f\left(M\left(t\right)\right)=c_{0}$,
so that $M\left(t\right)$ is constant. This fails to satisfy boundary
conditions on $M\left(t\right)$ and a minimizing solution would not
exist. A solution would occur only if $i\left(t\right)=0$, in which
case the Euler-Lagrange equation is trivially solved, with all pathways
yielding the same expenditure.}

Let us consider the influences on evolution of abatement rate $\dot{M}\left(t\right)$
in Eq. (\ref{eq:p14}), noting that $\frac{\ddot{M}\left(t\right)}{\dot{M}\left(t\right)}$
on the left describes its growth rate. 
\begin{itemize}
\item A positive interest rate $i\left(t\right)>0$ induces late abatement
by increasing the growth rate of $\dot{M}\left(t\right)$. 
\item Economic growth leads to future emissions growth and expansion of
abatement possibilities, through the term in $r_{BAU}\left(t\right)$.
This induces delayed abatement in the optimal solution, involving
larger growth rate of $\dot{M}\left(t\right)$ if the GDP growth rate
is higher. 
\item Without counting learning or damages, the horizontal position in the
MAC, relative to maximum abatement rate, has a particularly simple
form. Writing
\begin{equation}
\sigma\left(t\right)=\frac{\dot{M}\left(t\right)}{\dot{M}_{max}\left(t\right)}\label{eq:p15}
\end{equation}
which equals one when the annual emissions rate is exactly zero, its
growth rate is
\begin{equation}
\frac{1}{\sigma\left(t\right)}\frac{d\sigma\left(t\right)}{dt}=\frac{i\left(t\right)}{c_{2}}\label{eq:p16}
\end{equation}
\item If the interest rate is positive, the presence of an additive model
of learning with $f\left(M\left(t\right)\right)>0$ induces early
abatement by decreasing the growth rate of $\dot{M}\left(t\right)$. 
\item In contrast the multiplicative model of learning with $h'\left(M\left(t\right)\right)<0$
induces late abatement, regardless of the value of the interest rate.
\item Climate damages induce early abatement, since $d'\left(M\left(t\right)\right)<0$,
with higher initial abatement rate followed by smaller growth rate
of $\dot{M}\left(t\right)$. 
\end{itemize}

\subsection{Effects of the learning model}

The basic distinction pertaining to the endogenous learning model
is whether it appears as an additive or multiplicative contribution
in the abatement cost function. An additive model causes early abatement
in the presence of non-zero interest rate, whereas a multiplicative
model induces late abatement regardless of time-valuation.

As show in Appendix 4, the marginal abatement cost (MAC) corresponding
to our average cost model in Eq. (\ref{eq:p3}) is 
\begin{equation}
\beta\left(M,\dot{M}\right)=\left\{ c_{0}+\frac{\left(c_{2}+1\right)c_{1}}{\dot{M}_{max}^{c_{2}}}\dot{M}^{c_{2}}-f\left(M\right)\right\} h\left(M\right)\label{eq:p17}
\end{equation}
and the additive learning term $-f\left(M\left(t\right)\right)$,
being negative, reduces marginal cost by the same amount for all values
of abatement rate, thus effecting a downward shift in the MAC curve.
In contrast multiplication by $h\left(M\left(t\right)\right)$ pivots
the MAC downward, reducing it by the same factor for all values of
$\dot{M}$.

The effect of the additive model on growth rate of $\dot{M}\left(t\right)$
can be written as
\begin{equation}
c_{1}c_{2}\left(c_{2}+1\right)\left(\frac{\dot{M}\left(t\right)}{\dot{M}_{max}\left(t\right)}\right)^{c_{2}}\frac{\ddot{M}\left(t\right)}{\dot{M}\left(t\right)}=i\left(t\right)\frac{\beta\left(M\left(t\right),\dot{M}\left(t\right)\right)}{h\left(M\left(t\right)\right)}\label{eq:p18}
\end{equation}
where the effect of a positive $f\left(M\right)$ on the MAC has been
taken into account in Eq. (\ref{eq:p17}), and since it lowers the
MAC $\beta\left(M\left(t\right),\dot{M}\left(t\right)\right)$ its
effect is to moderate the growth rate of abatement without making
it negative. Generally it cannot induce higher abatement rates early
on. 

The multiplicative term appears on the right as 
\begin{equation}
-\frac{c_{1}c_{2}}{\dot{M}_{max}^{c_{2}}\left(t\right)}\dot{M}\left(t\right)^{c_{2}+1}\frac{h'\left(M\left(t\right)\right)}{h\left(M\left(t\right)\right)}\label{eq:p19}
\end{equation}
and, since $h'\left(M\right)<0$ the net contribution of the learning
model to the right side is positive, with its effect being to increase
emissions rate with time. This is opposite to the effect of the additive
model, furthermore it manifests independently of the interest rate.
Endogenous learning through the multiplicative term $h\left(N\right)$
is like discounting in time, reducing marginal costs by a common factor
for all values of $\dot{M}\left(t\right)$, with the difference that
in one case the reduction is a function of $M\left(t\right)$ whereas
in the other directly of $t$. Therefore it is hardly surprising that
such a learning model induces delayed abatement compared to the no-learning
case. 

As for the additive model, recall that its effect only occurs for
non-zero interest rate. The corresponding term $f\left(M\right)$
in Eq. (\ref{eq:p14}) is multiplied by $i\left(t\right)$, so the
effect vanishes for zero interest rate. This much is obvious from
the equation, but the underlying reasons are instructive and discussed
in Appendix 5. There it is shown that in the absence of discounting,
the variational problem reduces to minimizing a convex functional.
From Jensen's inequality this is solved by the expectation of the
abatement rate, which is in fact the constant-rate trajectory. 

\section{Applications}

\subsection{Evolution of the carbon tax}

How should the carbon tax evolve in order to induce the abatement
rates that would minimize the present value of costs? Since a carbon
tax of $P$ induces abatement activities whose marginal cost $\beta\left(M,\dot{M}\right)\leq P$,\footnote{Here we neglect transaction costs, option values, etc. that would
render this assumption inexact. } the evolution of the marginal abatement cost indicates this, and
from Eqs. (\ref{eq:p14}) and (\ref{eq:p17}), the carbon-tax growth
rate
\begin{multline}
\frac{dP\left(t\right)}{dt}=i\left(t\right)P\left(t\right)+\left\{ \gamma\left(\dot{M}\left(t\right),M\left(t\right)\right)\frac{h'\left(M\left(t\right)\right)}{h\left(M\left(t\right)\right)}-f'\left(M\left(t\right)\right)h\left(M\left(t\right)\right)\right\} \dot{M}\left(t\right)\\
+d'\left(M\left(t\right)\right)G\left(t\right)\label{eq:p19-1}
\end{multline}

comprises effects of the interest rate, endogenous learning, and economic
damages from climate change. In the absence of learning and damages,
the carbon tax growth $\frac{dP}{dt}=iP$ follows Hotelling's rule
regarding rent-extraction from nonrenewable resources (\citet{Hotelling1931}),
and parallels prior analyses (\citet{Dietz2019,Emmerling2019}), and
Eq. (\ref{eq:p19-1}) extends its validity to time-varying interest
rate. Endogenous learning reduces the growth rate, with $h'/h$ and
$-f'$ both being negative. Limiting damage costs in addition to meeting
the cumulative emissions constraint necessitates higher taxes earlier
on, and subsequent lower growth-rates of the carbon tax, and this
is quantified by $d'\left(M\right)<0$. 

\subsection{Cost of delaying abatement}

The IPCC, in its 2013 assessment, acknowledged the lessons of proportionality
between CO\textsubscript{2} induced global warming and cumulative
emissions for carbon budgets: ``The ratio of GMST {[}Global Mean
Surface Temperature{]} change to total cumulative anthropogenic carbon
emissions is relatively constant and independent of the scenario,
but is model dependent, as it is a function of the model cumulative
airborne fraction of carbon and the transient climate response. For
any given temperature target, higher emissions in earlier decades
therefore imply lower emissions by about the same amount later on''
(\citet{Stocker2013a}). The last statement might be interpreted to
mean that a delay in starting abatement can be compensated by more
stringent abatement in the future, but this would be expensive owing
to increasing marginal costs of abatement. 

What is the monetary cost of delaying abatement in the model, assuming
that the optimal emissions pathway is followed? Imagine that abatement
commences at $t=N$ years, where $0\leq N<T$. In the absence of endogenous
learning, the present value of the abatement cost is 
\begin{equation}
C_{total}=c_{1}\int_{N}^{T}e^{-I\left(t\right)}\sigma\left(t\right)^{c_{2}+1}\dot{M}_{max}\left(t\right)dt\label{eq:p21}
\end{equation}
assuming zero intercept $c_{0}=0$, and using Eqs. (\ref{eq:p2})
and (\ref{eq:p15}). Since $\dot{M}_{max}\left(t\right)$ is set equal
to the BAU emissions, which grows at $\theta r\left(t\right)$, the
optimal solution has $\sigma\left(t\right)$ growing at the rate $i\left(t\right)/c_{2}$,
and the initial abatement rate is constrained by the cumulative constraint
$\int_{N}^{T}\dot{M}\left(t\right)dt=M_{tot},$ integration yields
\begin{equation}
C_{total}=c_{1}\left(\frac{M_{tot}}{\dot{M}_{max}\left(0\right)}\right)^{c_{2}+1}\dot{M}_{max}\left(0\right)\frac{1}{\left\{ \int_{N}^{T}e^{I\left(t\right)/c_{2}+\theta R\left(t\right)}dt\right\} ^{c_{2}}}\label{eq:p22}
\end{equation}
as shown in Appendix 6. The last term increases with delayed starting
year $N$, corresponding to starting farther along the marginal cost
curve to meet the cumulative abatement constraint. Thus, a one-year
delay in mitigation leads to relative growth of total cost's present
value
\begin{equation}
\frac{1}{C_{total}}\frac{\partial C_{total}}{\partial N}=c_{2}\frac{e^{I\left(N\right)/c_{2}+\theta R\left(N\right)}}{\int_{N}^{T}e^{I\left(t\right)/c_{2}+\theta R\left(t\right)}dt}\label{eq:p23}
\end{equation}

and for the special case of constant rate of interest and GDP growth
at $i$ and $r$ respectively, this becomes
\begin{equation}
\frac{1}{C_{total}}\frac{\partial C_{total}}{\partial N}=c_{2}\left(\frac{i}{c_{2}}+\theta r\right)\frac{e^{\left(\frac{i}{c_{2}}+\theta r\right)N}}{e^{\left(\frac{i}{c_{2}}+\theta r\right)T}-e^{\left(\frac{i}{c_{2}}+\theta r\right)N}}\label{eq:p24}
\end{equation}
Our time-horizons in climate change mitigation typically specify end-of-century
goals, so we stipulate that $T$ is long and therefore:
\begin{itemize}
\item if $N$ is small, $\frac{1}{C_{total}}\frac{\partial C_{total}}{\partial N}\approx\frac{c_{2}\left(\frac{i}{c_{2}}+\theta r\right)}{e^{\left(\frac{i}{c_{2}}+\theta r\right)T}}$
\item whereas if $N$ is large, approaching $T$, $\frac{1}{C_{total}}\frac{\partial C_{total}}{\partial N}\approx\frac{c_{2}}{T-N}$
\end{itemize}
Evidently, delay in the start of mitigation (i.e. starting onto the
optimal abatement trajectory) is costly in present-value terms because
of increasing marginal costs of abatement. Such a delay entails higher
abatement rates in the future, leading to higher costs. 

The growth rate of the present value of abatement costs due to each
year of delay in mitigation is independent of the abatement level,
but depends on the time-horizon and the shape of the MAC. Delaying
abatement is more costly if $c_{2}$ is higher, because rapid rise
in the MAC requires higher initial abatement rates to limit future
growth in costs. In contrast, assuming small $N$, delaying abatement
is expensive but less so for high GDP growth or large interest rate,
because these conditions lead to smaller starting abatement rates.
Delay becomes more costly if the time-horizon is short. As the starting
year for mitigation $N$ approaches $T-1$, the present-value of abatement
cost grows at an increasing rate approaching $c_{2}$. 

\subsection{Initial carbon tax}

How does such a delay in starting abatement influence the initial
carbon tax necessary? The cumulative abatement constraint, without
counting learning or damages, becomes (Appendix 6)
\begin{equation}
M_{tot}=\int_{N}^{T}\dot{M}\left(t\right)dt=\dot{M}\left(N\right)e^{-I\left(N\right)/c_{2}}e^{-\theta R\left(N\right)}\int_{N}^{T}e^{I\left(t\right)/c_{2}+\theta R\left(t\right)}dt\label{eq:p25}
\end{equation}
and using $\dot{M}_{max}\left(N\right)=\dot{M}_{max}\left(0\right)e^{\theta R\left(N\right)}$,
the position along the MAC curve $\sigma\left(N\right)=\dot{M}\left(N\right)/\dot{M}_{max}\left(N\right)$
at the start of abatement is
\begin{equation}
\sigma\left(N\right)=\frac{M_{tot}}{\dot{M}_{max}\left(0\right)}\frac{e^{I\left(N\right)/c_{2}}}{\int_{N}^{T}e^{I\left(t\right)/c_{2}+\theta R\left(t\right)}dt}\label{eq:p26}
\end{equation}
which grows with $N$, so meeting a cumulative abatement following
a delay requires one to start farther along the curve. This gives
initial marginal cost, from Eq. (\ref{eq:p17}), equal to $\beta\left(N\right)=c_{1}\left(c_{2}+1\right)\sigma\left(N\right)^{c_{2}}$
so the starting carbon tax must be
\begin{equation}
P\left(N\right)=c_{1}\left(c_{2}+1\right)\left(\frac{M_{tot}}{\dot{M}_{max}\left(0\right)}\right)^{c_{2}}\frac{e^{I\left(N\right)}}{\left\{ \int_{N}^{T}e^{I\left(t\right)/c_{2}+\theta R\left(t\right)}dt\right\} ^{c_{2}}}\label{eq:p27}
\end{equation}
As before, we consider effects of delayed abatement
\begin{equation}
\frac{1}{P}\frac{\partial P}{\partial N}=i\left(N\right)+c_{2}\frac{e^{I\left(N\right)/c_{2}+\theta R\left(N\right)}}{\int_{N}^{T}e^{I\left(t\right)/c_{2}+\theta R\left(t\right)}dt}\label{eq:p28}
\end{equation}
which is similar to Eq. (\ref{eq:p23}), but with addition of Hotelling's
growth-rate $i\left(t\right)$. For constant $r$ and $i$, the corresponding
limiting cases for long time-horizon $T$ are:
\begin{itemize}
\item $N$ small, $\frac{1}{P}\frac{\partial P}{\partial N}\approx i+\frac{c_{2}\left(\frac{i}{c_{2}}+\theta r\right)}{e^{\left(\frac{i}{c_{2}}+\theta r\right)T}}$
\item whereas $N$ large, approaching $T$, $\frac{1}{P}\frac{\partial P}{\partial N}\approx i+\frac{c_{2}}{T-N}$
\end{itemize}
and the growth of the tax has a contribution from the Hotelling rate
$i\left(t\right)$. Additionally each year's delay in implementation
gives rise to a further increase, owing to increasing marginal costs. 

\subsection{Economics of avoiding overshoot}

Given proportionality between global warming and cumulative emissions,
overshoot in CO\textsubscript{2}'s warming contribution generally
corresponds to net negative emissions. Scenarios avoiding overshoot
therefore must have positive but declining emissions for $t<T$, so
that $\sigma\left(t\right)<1$. The optimal pathways generally have
$\sigma\left(t\right)$ growing in time, so we must only consider
the constraint $\sigma\left(T\right)<1$ for avoiding an overshoot.
Since the position along the abatement curve grows, without considering
learning or damages, at $\sigma\left(T\right)=\sigma\left(N\right)e^{\int_{N}^{T}i\left(t\right)/c_{2}dt}$,
we obtain from Eq. (\ref{eq:p62}) in Appendix 6
\begin{equation}
\sigma\left(T\right)=\frac{M_{tot}}{\dot{M}_{max}\left(0\right)}\frac{e^{I\left(T\right)/c_{2}}}{\int_{N}^{T}e^{I\left(t\right)/c_{2}+\theta R\left(t\right)}dt}\label{eq:p29}
\end{equation}
Furthermore, in the absence of net negative emissions, future global
warming is roughly proportional to future cumulative emissions via
the TCRE $\alpha$, so that $\triangle T_{fut}=\alpha\left\{ E_{BAU,fut}\left(T\right)-M_{tot}\right\} $.
Future cumulative emissions under BAU is simply $E_{BAU,fut}\left(T\right)=\dot{M}_{max}\left(0\right)\int_{0}^{T}e^{\theta R\left(t\right)}dt$.
Hence 
\begin{equation}
\frac{M_{tot}}{\dot{M}_{max}\left(0\right)}=\int_{0}^{T}e^{\theta R\left(t\right)}dt-\frac{\triangle T_{fut}}{\alpha\dot{M}_{max}\left(0\right)}\label{eq:p30}
\end{equation}
and substituting in Eq. (\ref{eq:p29}), the constraint $\sigma\left(T\right)<1$
for avoiding overshoot is
\begin{equation}
\frac{\triangle T_{fut}}{\alpha\dot{M}_{max}\left(0\right)}>\int_{0}^{T}e^{\theta R\left(t\right)}dt-\frac{\int_{N}^{T}e^{I\left(t\right)/c_{2}+\theta R\left(t\right)}dt}{e^{I\left(T\right)/c_{2}}}=\frac{\triangle T_{fut}^{*}}{\alpha\dot{M}_{max}\left(0\right)}\label{eq:p31}
\end{equation}
providing a lower bound $\triangle T_{fut}^{*}$ on future global
warming from CO\textsubscript{2} attainable without overshoot, assuming
an optimal pathway is being followed. Moreover, each year of delay
in starting mitigation causes increase in this minimum attainable
temperature
\begin{equation}
\frac{\partial\triangle T_{fut}^{*}}{\partial N}=\alpha\dot{M}_{max}\left(0\right)e^{-\int_{N}^{T}\frac{i\left(t\right)}{c_{2}}dt+\theta R\left(N\right)}>0\label{eq:p32}
\end{equation}
with the marginal effects of each subsequent year of delay rising
as $\partial^{2}\triangle T_{fut}^{*}/\partial N^{2}=\left\{ \frac{i}{c_{2}}\left(N\right)+\theta r\left(N\right)\right\} \frac{\partial\triangle T_{fut}^{*}}{\partial N}>0$,
as long as $\frac{i}{c_{2}}+\theta r>0$. Hence, each year of delay
makes it increasingly difficult to meet global warming goals without
overshoot. For constant $i$ and $r$
\begin{equation}
\triangle T_{fut}^{*}=\alpha\dot{M}_{max}\left(0\right)\left\{ \frac{e^{\theta rT}-1}{\theta r}-\frac{e^{\theta rT}}{\frac{i}{c_{2}}+\theta r}\left(1-e^{-\left(\frac{i}{c_{2}}+\theta r\right)\left(T-N\right)}\right)\right\} \label{eq:p33}
\end{equation}
and, as before, we consider two cases, assuming long $T$:
\begin{itemize}
\item $N$ is small, $\triangle T_{fut}^{*}\approx\alpha\dot{M}_{max}\left(0\right)\frac{e^{\theta rT}}{\theta r}\frac{1}{1+\frac{\theta r}{i/c_{2}}}=\alpha\frac{\dot{M}_{max}\left(T\right)}{\theta r}\frac{1}{1+\frac{\theta r}{i/c_{2}}}$
\item $N$ is large, making $T-N$ is small, so that $\triangle T_{fut}^{*}\approx\alpha\dot{M}_{max}\left(T\right)\left\{ \frac{1}{\theta r}-\left(T-N\right)\right\} $
\end{itemize}
These results have a simple interpretation in terms of cumulative
emissions under BAU, which is roughly $\dot{M}_{max}\left(T\right)/\theta r$
for long time-horizons. This means that, if there is an early start
to global abatement, a fraction $\theta r/\left(\frac{i}{c_{2}}+\theta r\right)$
of BAU emissions is eliminated in optimizing scenarios without overshoot.
Lower rates of interest favor eliminating a larger fraction of BAU
emissions, by giving rise to higher abatement rates early on. Of course,
the actual temperature threshold obtained increases proportionally
to the TCRE, whose magnitude is uncertain. In case of a late start
to abatement, in the limit when $N$ approaches $T$ the results trivially
show that the threshold temperature is practically governed by BAU
emissions. 

\section{Computational examples}

\subsection{Marginal abatement cost and economic growth assumptions}

Our optimal pathways are based on the Euler-Lagrange Eq. (\ref{eq:p14}),
whose derivation fixes endpoint variations $\delta M\left(0\right)=\delta M\left(T\right)=0$,
and hence $M\left(0\right)$ and $M\left(t\right)$ are fixed, but
leaving unrestrained the initial slope $\dot{M}\left(0\right)$. Therefore
the present value of the emissions-rate $\dot{E}\left(0\right)$ is
not specified by the model, giving rise to a discontinuity in the
optimal abatement trajectories at the start of abatement. We adopt
some parameters from the DICE model (\citet{Nordhaus2013}), such
as the shape of the marginal abatement cost (Figure 1) in Eq. (\ref{eq:p46}),
with zero intercept and maximum cost of $550\$$ per ton (or billion
$\$$ per Gton) of CO\textsubscript{2} (at 2015 values). At higher
values of abatement, where $\sigma\left(t\right)>1$ the same formula
is extended. As noted earlier, the additive model of learning effects
a uniform downward shift, whereas a multiplicative model reduces cost
by a constant factor (Figure 1b). 
\begin{figure}
\includegraphics[scale=0.7]{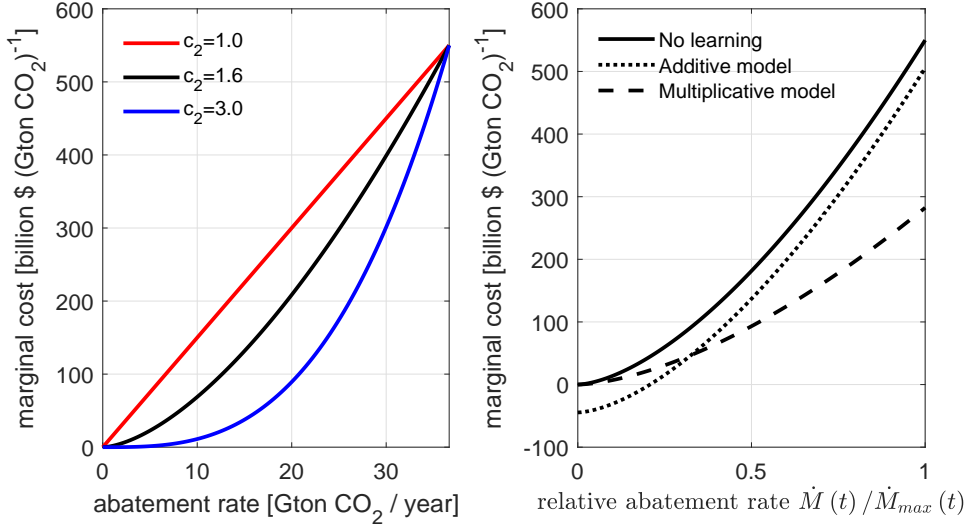}\caption{Marginal abatement cost curves with different exponents and the effects
of additive and multiplicative models of endogenous learning for $c_{2}=1.6$.}
\end{figure}

The optimal abatement rate grows with time due to combined effects
of economic growth and the interest rate. Taking income elasticity
of global CO\textsubscript{2} emissions as $\theta=0.75$ , emissions
intensity under BAU evolves as $\frac{d\mu}{dt}\cong-\left(1-\theta\right)r\mu$
and, with $r=4$\%, this becomes $\frac{d\mu}{dt}\approx-0.01\mu$
, mimicking the rate of exogenous reduction in emissions intensity
within DICE (\citet{Nordhaus2013}). Then BAU emissions grow at $r_{BAU}\left(t\right)=0.75r\left(t\right)$,
with the parameter $\dot{M}_{max}\left(t\right)$ also rising at this
rate. For the simulations below, $r\left(t\right)$ decreases exponentially
from the present value $r\left(0\right)=4$\% with an e-folding time
of $\tau_{r}=40$ years, so that $R\left(t\right)=\tau_{r}r\left(0\right)\left(1-e^{-t/\tau_{r}}\right)$
and $\dot{E}_{BAU}\left(t\right)=\dot{E}_{BAU}\left(0\right)e^{\theta\tau_{r}r\left(0\right)\left(1-e^{-t/\tau_{r}}\right)}$
so cumulative emissions under BAU are estimated numerically. 

\subsection{Optimal abatement pathways and carbon tax growth rates}

The optimal abatement rate in the absence of learning or damages grows
as $\frac{1}{\dot{M}\left(t\right)}\frac{d\dot{M}\left(t\right)}{dt}=\frac{i\left(t\right)}{c_{2}}+\theta r\left(t\right)$,
and is integrated for the abatement rate in Figure 2a. GDP growth
leads to expansion of abatement possibilities and rise in the abatement
rate with time, but this effect is decreasing with declining economic
growth, as seen in the declining slope on the logarithmic scale. All
the above effects are simply absorbed in evolution of the carbon tax,
which increases at constant Hotelling's rate $i$ in the absence of
learning or damages (Eq. (\ref{eq:p19-1}) and Figure 2d). Here the
interest rate is assumed constant in time, however this restriction
is not generally posed. 

The main feature of the endogenous learning model, as noted earlier,
is whether it appears through an additive or multiplicative term.
Empirical studies indicate a multiplicative model, having power-law
structure $h\left(M\right)=M^{-b}$, with $b>0$ (\citet{Rubin2004,Lindman2012}),
adapted here to $h\left(M\right)=\left(M\left(t\right)/M\left(t=0\right)\right)^{-b}$
with $M\left(t=0\right)>0$ in order that $h\left(0\right)=1$. This
has constant relative reductions of marginal cost, by $\left(\frac{1}{2}\right)^{b}$,
per doubling of cumulative abatement. We also try exponential model
$h\left(M\right)=e^{-M/M_{h}}$, with $M\left(t=0\right)\ll M_{h}$
so that $h\left(0\right)\cong1$, giving rise to proportional reductions
in marginal cost per unit increase in cumulative abatement, with constant
$\partial\beta/\partial M=-\beta/M_{h}$. Such a choice depends on
whether proportional cost reductions arise from doublings or unit
increases of cumulative abatement. In both cases, the multiplicative
model of learning has delayed abatement, compared to the no-learning
case, with initially lower abatement rates later exceeding the no-learning
case (Figure 2a). The exponential and power-law models do differ in
important ways, with the latter's influence declining with time, whereas
the exponential model makes a growing difference with time. The power-law
model depends on doublings of cumulative abatement, which become infrequent
with time, whereas the exponential model depends on constant changes
in cumulative abatement which occur more rapidly as abatement proceeds. 

Emissions (Figure 2b) for these scenarios display the aforementioned
discontinuity at the start of abatement. Global warming (Figure 2c)
shows, for the more stringent abatement case, overshoot of global
warming owing to the requirement of negative emissions for meeting
the cumulative abatement goal. The overshoot is larger with the exponential
learning model, owing to slower abatement early on. The carbon tax
(Figure 2d) grows more slowly with learning, as shown earlier, with
its magnitude being initially lower compared to the no-learning case
for these multiplicative models, because initial abatement rates are
reduced in this case, giving rise to smaller carbon taxes throughout.
The exponential model has learning progressing rapidly, and thus for
the stringent mitigation cases the departure from Hotelling's rule
is greater because the relative effect of learning on the growth rate
is, from Eq. (\ref{eq:p19-1}), $-\left(c_{2}+1\right)\frac{\dot{M}\left(t\right)}{i\left(t\right)M_{h}}$,
whose magnitude increases with the abatement rate. 

\begin{figure}
\includegraphics[scale=0.75]{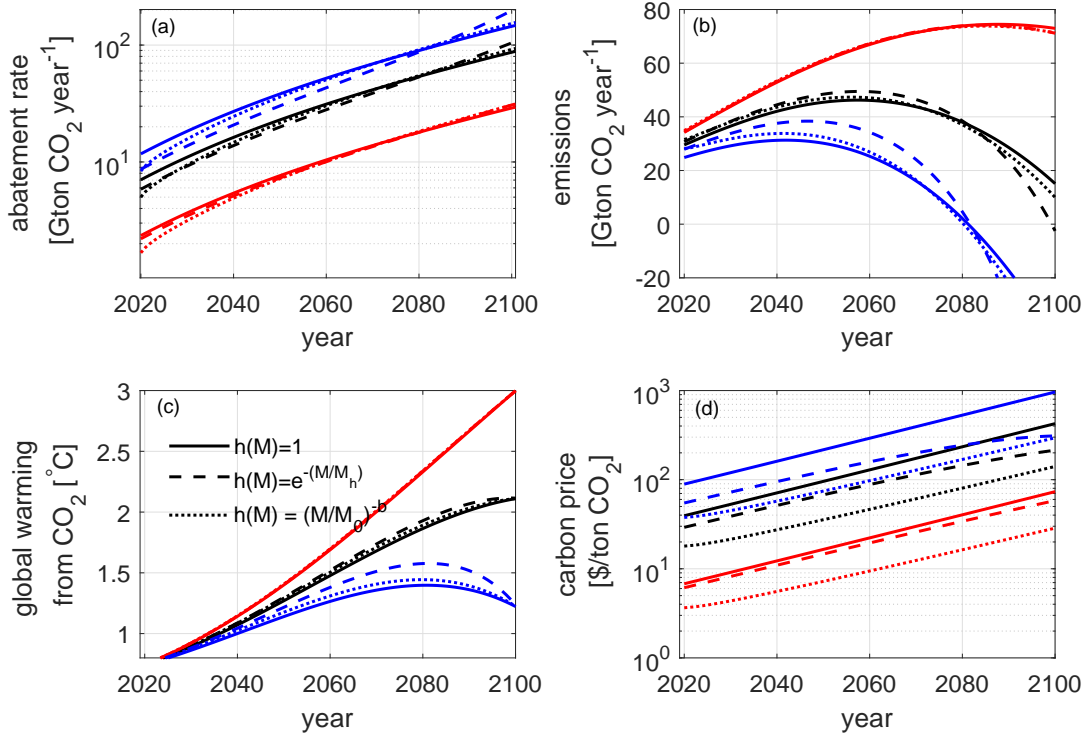}

\caption{Effects of learning models with a multiplicative term: (a): Optimal
abatement rates solving Eq. (\ref{eq:p14}); (b): Emissions of CO\protect\textsubscript{2};
(c) Global warming contribution of CO\protect\textsubscript{2}; (d)
Evolution of global carbon tax. The interest rate is $i=0.03$. Red,
black and blue curves correspond to cumulative abatement goals of
$M_{tot}=1000$ ,$M_{tot}=3000$, and $M_{tot}=5000$ Gton CO\protect\textsubscript{2}
respectively. The carbon tax grows at the rate of approximately $0.03$
per year in the no-learning case as expected from Eq. (\ref{eq:p19-1}),
with estimated $95$-percent confidence interval of $\left[0.0297,0.0298\right]$
for the mean growth rate in numerical calculations. For the exponential
model, the corresponding $95$-percent confidence intervals are $\left[0.0281,0.0283\right]$,
$\left[0.0249,0.0256\right]$, and $\left[0.0219,0.0231\right]$ for
$M_{tot}$ of $1000$, $3000$, and $5000$ Gton respectively. For
the power-law model these are the same $\left[0.0263,0.0267\right]$
for all the different values of $M_{tot}$.}
\end{figure}
An additive model of learning, such as $f\left(M\right)=c_{f}M$,
has higher abatement rate early on compared to the no-learning case
(Figure 3a). Overshoot is marginally lower in the stringent abatement
case (Figure 3c), owing to lower emissions early on (Figure 3b). This
comes from higher carbon taxes early on (Figure 3d), which become
lower than the no-learning case only much later. Generally, additive
models of learning lead to higher average carbon taxes than multiplicative
models of learning. Since learning is proportional to cumulative abatement,
the largest departures from Hotelling's rule occur for the less stringent
abatement scenarios, because from Eq. (\ref{eq:p19-1}) its relative
effect is $-\frac{c_{f}}{i\left(t\right)\left\{ \beta\left(M\left(t\right),\dot{M}\left(t\right)\right)/\dot{M}\left(t\right)\right\} }$
which is decreasing in $\dot{M}$ as the marginal cost curve is strictly
convex. 

\begin{figure}
\includegraphics[scale=0.75]{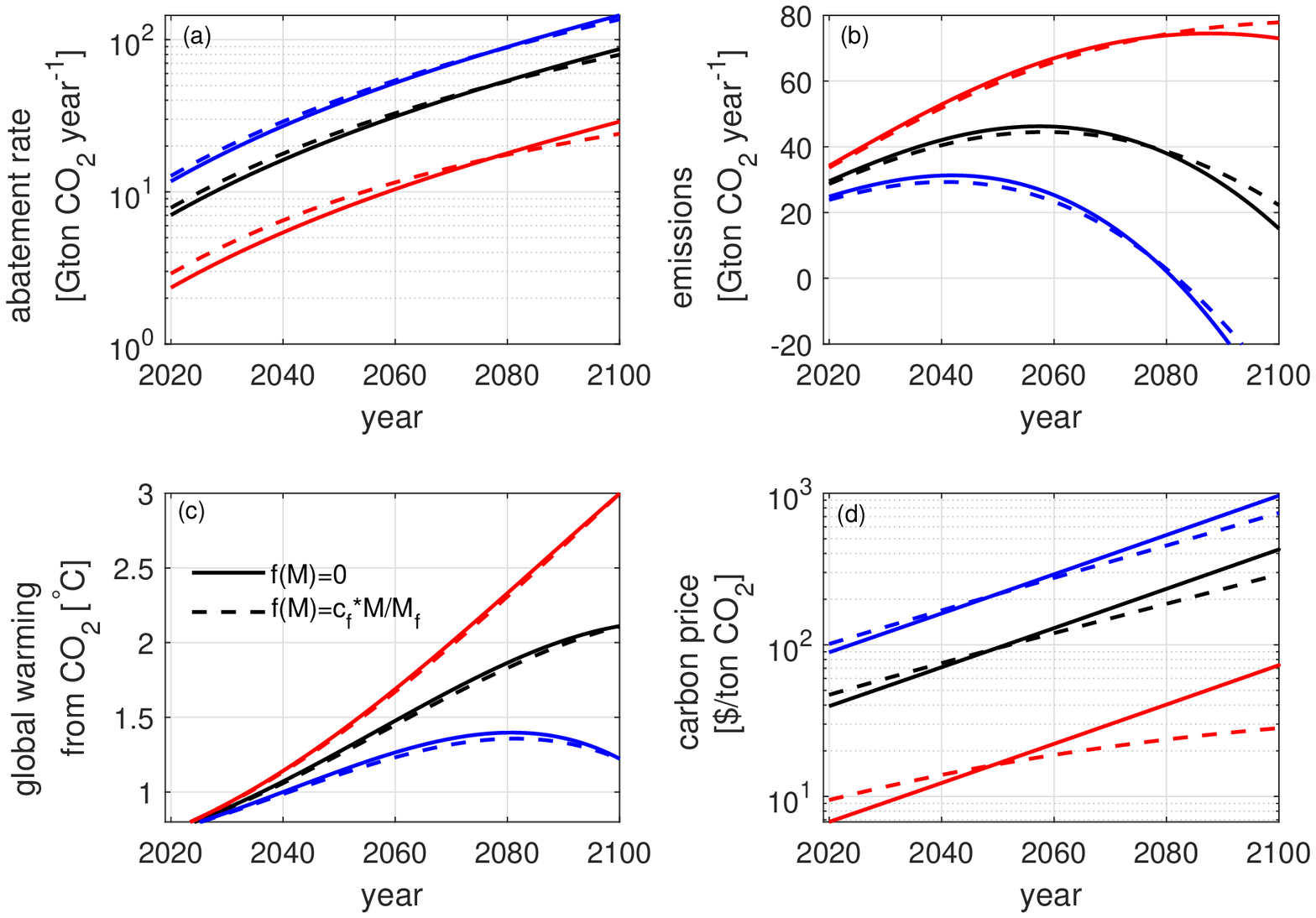}

\caption{Effects of learning models with an additive term: (a): Optimal abatement
rates solving Eq. (\ref{eq:p14}); (b): Emissions of CO\protect\textsubscript{2};
(c) Global warming contribution of CO\protect\textsubscript{2}; (d)
Evolution of global carbon tax. The interest rate is $i=0.03$. Red,
black and blue curves correspond to cumulative abatement goals of
$M_{tot}=1000$ ,$M_{tot}=3000$, and $M_{tot}=5000$ Gton CO\protect\textsubscript{2}
respectively. The carbon tax grows at the rate of approximately $0.03$
per year in the no-learning case as before. For the additive learning
model shown here, the estimated $95$-percent confidence intervals
of the mean growth rate are $\left[0.0130,0.0139\right]$, $\left[0.0226,0.0228\right]$,
and $\left[0.0247,0.0248\right]$ for $M_{tot}$ of $1000$, $3000$,
and $5000$ Gton respectively. }

\end{figure}

\subsection{Effects of climate damage models}

While typically, climate damages are treated as a function of global
warming (\citet{Nordhaus1993,Dietz2019}), for which there is considerable
evidence (\citet{Burke2015}), and hence modeled as a function of
cumulative abatement in our framework, in general one might also expect
other influences. In particular, one might expect terms $\dot{M}\left(t\right)$,
and $\int_{0}^{t}M\left(s\right)ds$ in conjunction with stochastic
terms to play a role, to account for effects of rate of global warming,
global sea-level rise whose rate of change also depends on the present
global warming (\citet{Rahmstorf2007,Vermeer2009}),\footnote{With the rate of global mean sea-level rise written as $\frac{dS}{dt}\approx s_{1}\triangle T\left(t\right)+s_{2}\frac{d\triangle T\left(t\right)}{dt}$
(\citet{Vermeer2009}), its impacts would depend on $\int_{-T_{PI}}^{t}\triangle T\left(s\right)ds+\triangle T\left(t\right)$,
and hence on integral of cumulative abatement in addition to the cumulative
abatement itself.} and lastly internal weather and climate variability. Despite its
importance, optimization relative to the time-integral of $M\left(t\right)$
is not admitted by our present framework although such an extension
is definitely possible,\footnote{The functional that is minimized to yield the Euler-Lagrange equation
is assumed here to depend only on $M$ and $\dot{M}$, but not $\int M$,
but Euler also treated the more general case that included extremization
relative to integrals (\citet{Fraser1992}). } and hence we retain a model in $M\left(t\right)$ alone while noting
its limitations, including the omission of slow and delayed effects
of sea-level rise whose consideration can influence abatement pathways. 

A model commonly adopted is the power-law function of global warming
(e.g. \citet{Nordhaus2013}), written here as $d\left(M\left(t\right)\right)=\frac{d_{0}}{T_{0}^{d_{1}}}\left\{ \alpha\left(E_{BAU}\left(t\right)-M\left(t\right)\right)\right\} ^{d_{1}}$
, for which damage fraction $d\left(t\right)$ grows $2^{d_{1}}$
for each doubling of warming $\triangle T$, which occur more frequently
at smaller values (Figure 4). An alternative is logistic $d\left(M\left(t\right)\right)=\frac{1}{1+\frac{1}{d_{2}}e^{-\left(E_{BAU}\left(t\right)-M\left(t\right)\right)/E_{D}}}$,
approximating the exponential for small $M\left(t\right)$, and $1$
for large values. Both models possess two free parameters, estimated
from specifying a $5$\% GDP damage fraction for $\Delta T=2.5$ K,
growing to $20$\% for $\Delta T=5$ K. Owing to this constraint,
for $\Delta T$ of a few degrees, the two models are similar (Figure
4), but the inset shows that the power-law is more sensitive. 

Including the minimization of damage costs leads to early abatement,
compared to simulations that do not also include such costs, with
higher initial abatement rates (Figure 5a) leading to somewhat delayed
negative emissions (Figure 5b) and lowering of the overshoot in temperature
(Figure 5c). The power-law model is more consequential, since the
damages are more sensitive to temperature change, and this leads to
higher starting carbon prices that grow at smaller rates compared
to the simulations where damage costs are not counted. The effect
is larger in scenarios with little abatement, because damage costs
are greater. 

\begin{figure}
\includegraphics[scale=0.75]{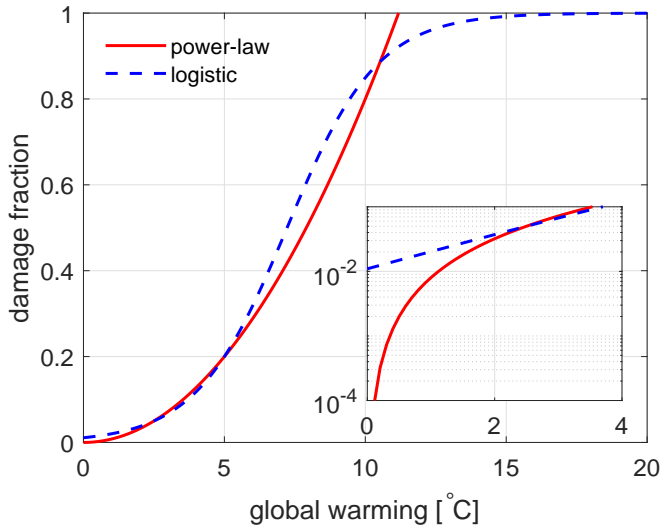}

\caption{Power law and logistic models of the damage fraction as a function
of global warming. For the range of global warming shown here, the
logistic model can be approximated by the exponential as it follows
a straight line on the logarithmic scale (inset). Damages in the power-law
model grow more rapidly over the temperature range of interest. }

\end{figure}

\pagebreak{}

\begin{figure}
\includegraphics[scale=0.85]{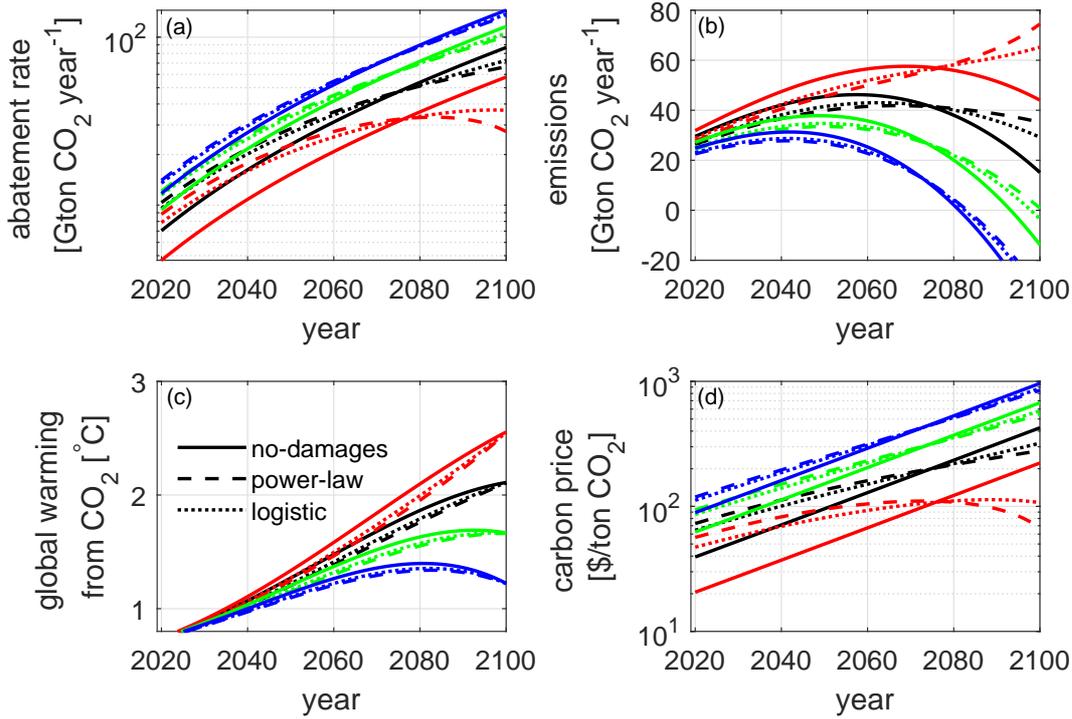}

\caption{Effects of damage models: (a): Optimal abatement rates solving Eq.
(\ref{eq:p14}); (b): Emissions of CO\protect\textsubscript{2}; (c)
Global warming contribution of CO\protect\textsubscript{2}; (d) Evolution
of global carbon tax. The interest rate is $i=0.03$. Red, black,
green and blue curves describe cumulative abatement goals of $M_{tot}=2000$
, $3000$, $4000$ and $5000$ Gton CO\protect\textsubscript{2} respectively.
The inclusion of damages leads to early abatement for the same cumulative
constraint. The power-law model leads to somewhat larger abatement
rates than the logistic model. Damages have a larger impact on the
abatement pathway and carbon taxes for the less stringent abatement
scenarios. }
\end{figure}

Multiplicative learning and damages have countervailing effects on
the abatement rate, and their relative effects for the exponential
and power-law models of learning and damages respectively are, from
Eq. (\ref{eq:p14}), equal to $\frac{c_{2}}{d_{1}}\frac{A\left(t\right)/M_{h}}{D\left(t\right)/E\left(t\right)}$,
where $A\left(t\right)=\gamma\left(t\right)\dot{M}\left(t\right)$
is the annual abatement cost from Eq. (\ref{eq:p9}) and $D\left(t\right)$
is damage cost. Damages are convex in global warming, and hence in
cumulative emissions $E\left(t\right)$. This makes the ratio of damages
per cumulative emissions in the denominator higher in scenarios with
high cumulative emissions, for which corresponding abatement costs
are lower. Thus we can expect the modeling of damage costs to play
a greater role in scenarios with weak abatement, and contrariwise
with stringent abatement. Similarly, while both learning and damages
lower the carbon tax growth rate, their relative importance for this
growth rate depends on the factor $\frac{1}{d_{1}}\frac{A\left(t\right)/M_{h}}{D\left(t\right)/E\left(t\right)}$.
The carbon tax growth rate depends more strongly on the uncertain
damage model in low-abatement scenarios, and on the learning model
in high-abatement scenarios (Figure 6). The axes are damage fraction
at $\triangle T=2.5$ K, fixing $d_{1}=2$, and learning parameter
$M_{h}$, with small and large values, in the respective limiting
cases, corresponding to learning and damages not being counted. 

\begin{figure}
\includegraphics[scale=0.7]{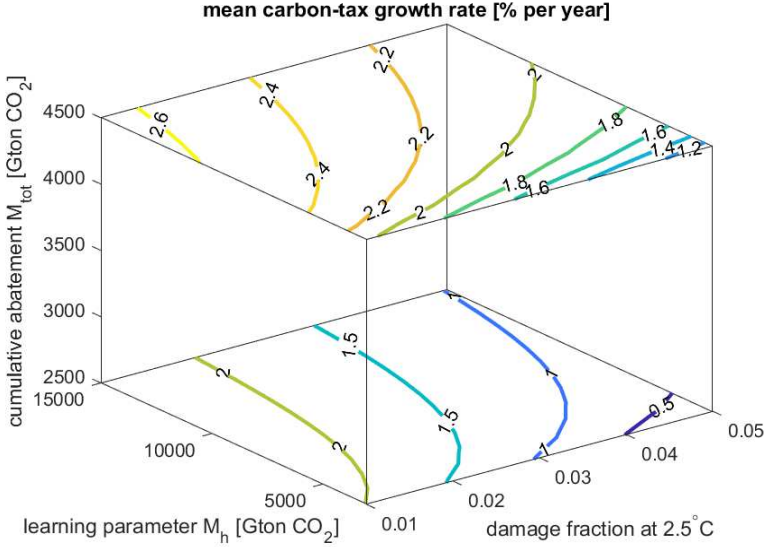}

\caption{Contours of the carbon tax growth rate, (in percent per year) as a
function of the damage fraction for global warming of 2.5 degrees
C and the learning parameter of the exponential model, for two different
cumulative abatement goals. The growth rate is always smaller than
the interest rate, which is taken to be $0.03$ .}
\end{figure}

\subsection{Initial carbon tax and cost of delaying abatement}

We derived an expression for initial carbon tax in Eq. (\ref{eq:p27}),
assuming that learning and damages are not counted in the optimal
pathway. This is graphed for constant GDP growth and interest rates,
for abatement starting at the present year with $N=0$. Substituting
from Eq. (\ref{eq:p30}) the initial tax is
\begin{equation}
P\left(0\right)=c_{1}\left(c_{2}+1\right)\left\{ \int_{0}^{T}e^{\theta R\left(t\right)}dt-\frac{\Delta T_{final}-\Delta T\left(0\right)}{\alpha\dot{M}_{max}\left(0\right)}\right\} \frac{1}{\left\{ \int_{0}^{T}e^{I\left(t\right)/c_{2}+\theta R\left(t\right)}dt\right\} ^{c_{2}}}
\end{equation}
where $\triangle T_{final}$ is global warming from preindustrial
times at the end of the time-horizon and $\Delta T\left(0\right)$
is the present global warming from past cumulative emissions of CO\textsubscript{2}.
Contour plots (Figure 7) show that, in addition to rapid GDP growth
and more stringent goals involving smaller $\Delta T_{final}$, the
initial tax is very sensitive to the interest rate. Low interest rates
give rise to larger abatement rates early on (compared to high interest
rates) in the optimal pathways, which must correspond to a higher
initial carbon tax. 

\begin{figure}
\includegraphics[scale=0.6]{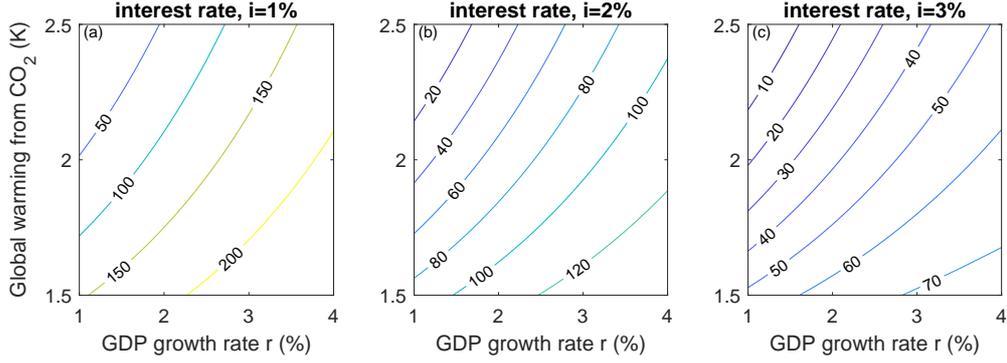}

\caption{Initial value of the carbon tax (\$ / ton CO\protect\textsubscript{2}),
assuming that abatement begins at the present, i.e. $N=0$. The carbon
tax is shown as contours of GDP growth rate and the total global warming
from CO\protect\textsubscript{2} $\Delta T_{final}$, for different
values of risk-free interest rate in each panel. Calculations assume
mean TCRE, of $1.65$ K / Tton C. The initial carbon tax is higher
for lower interest rates, high rates of GDP growth, and stringent
abatement goals. }
\end{figure}

An abatement delay by one year, postponing the onset of a carbon tax,
increases the required initial value of the tax. The relative increase,
shown in Eq. (\ref{eq:p28}), has two contributions: the annual growth
rate of the carbon tax by the Hotelling rule, and a second term describing
the increasing average costs of abatement borne due to the higher
abatement rates necessary for meeting the cumulative abatement goal.
The second contribution grows identically as the present value of
abatement cost for each year of delay, as comparison of Eqs. (\ref{eq:p24})
and (\ref{eq:p28}) showed. The reason is straightforward: the present
value (at $t=0$) of the average carbon tax (between years $t=N$
and $t=T)$ grows at this rate, for each year of delay, and with average
cost of mitigation being proportional to the marginal cost,\footnote{There is a constant factor of proportionality $1/\left(1+c_{2}\right)$
relating average to marginal costs when the intercept of the MAC curve
is zero.} the present value of the abatement cost also grows at this rate. 

This growth rate, from Eq. (\ref{eq:p24}), is graphed versus the
year of starting abatement $N$ (Figure 8), for different GDP growth
rates and interest rates. Its value increases for each year of delay,
approaching $c_{2}$ as $N$ approaches $T$. Lower interest rates,
by inducing higher rates of abatement early on, increasing the costs
of delaying abatement by one year. Higher GDP growth rates give rise
to larger growth in abatement rates with time, as abatement options
grow with BAU emissions. Hence rapid economic growth has the effect
of lowering the costs of delaying abatement, in relative terms. 

\begin{figure}
\includegraphics[scale=0.6]{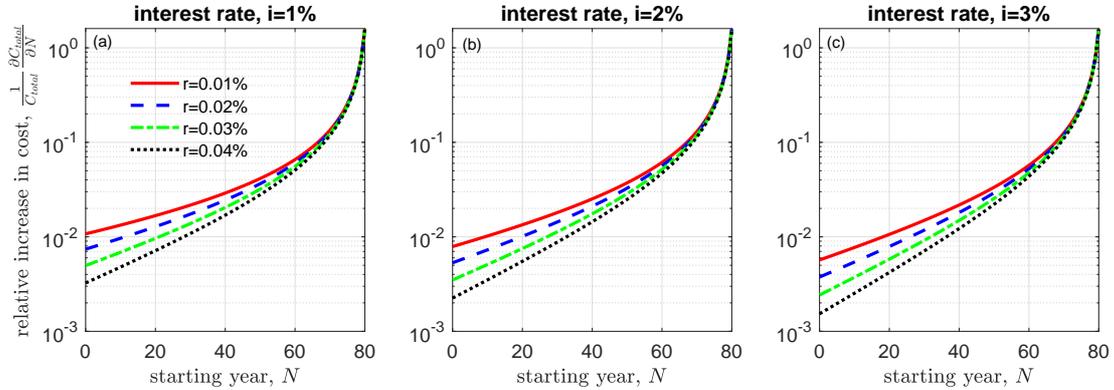}

\caption{Growth rate of the present value of cost of delaying abatement by
one year, in Eq. (\ref{eq:p24}), for different starting years $N$,
GDP growth rates, and interest rates. Growth rate is larger for lower
interest rate and GDP growth rates, and increases rapidly for each
year of delay. }
\end{figure}

\subsection{Minimum global warming without overshoot}

Our optimal pathways in Eq. (\ref{eq:p14}) do not specify the avoidance
of temperature overshoot, but this distinction is important. Therefore
we examined the threshold (i.e. minimum) global warming goal that
can be met without overshoot, in Eq. (\ref{eq:p33}), based on the
requirement of net positive emissions throughout, i.e. $\sigma\left(t\right)<1$
for all $t<T$, and reaching zero emissions only at $t=T$. Figure
9 shows the resulting threshold (i.e. minimum) CO\textsubscript{2}-induced
global warming attainable in 2100, given the history of CO\textsubscript{2}
emissions and mean TCRE, and through the optimal abatement pathways
modeled here. The starting year of abatement and the interest rate
are important constraints, in addition to GDP growth, because large
increases in abatement rate would pose increased abatement costs in
the future. As a result, if the start of abatement is considerably
delayed, options for substantially limiting CO\textsubscript{2}-induced
global warming become increasingly costly. Each year of delay leads
to an increase in the threshold global warming by an increasing amount,
because the abatement rate grows at a positive rate given by $\frac{i}{c_{2}}+\theta r$. 

Since optimal pathways can abate only a fraction of future BAU emissions,
rapid economic growth leads to an increase in the threshold. For example
for long time-horizons with an early start to abatement, roughly $\theta r/\left(\frac{i}{c_{2}}+\theta r\right)$
of BAU emissions is abated by the optimal pathway. Larger interest
rates lower the fraction of BAU emissions abated, because they lead
to smaller abatement rates early on, thereby raising the temperature
threshold. For example, with an interest rate of $3$ \% it does not
appear possible to meet a $1.5$ K global warming goal, even at mean
TCRE, let alone the higher estimates, unless future GDP growth were
to be practically negligible. In contrast if interest rates are closer
to $1$ \%, then in principle the more stringent global warming goals
can be met even with rapid GDP growth, but this corresponds to a high
carbon tax early on. 

\begin{figure}
\includegraphics[scale=0.7]{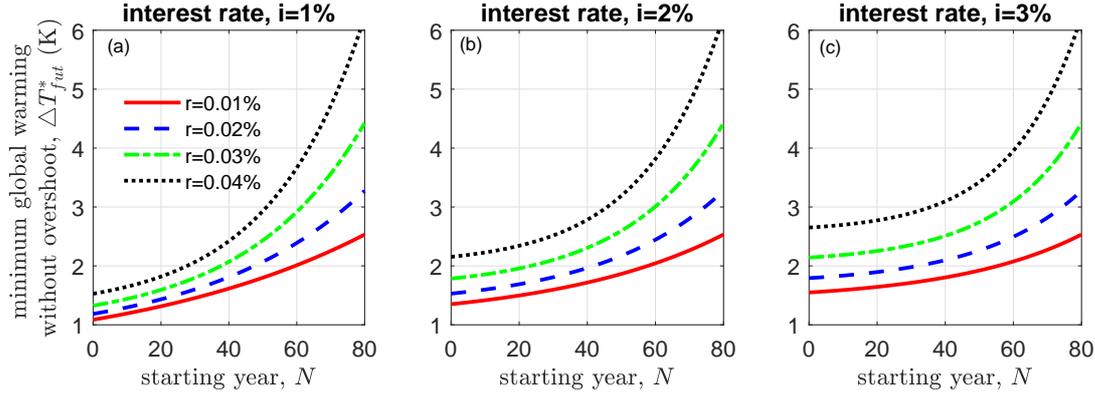}

\caption{The minimum CO\protect\textsubscript{2}-induced warming threshold
in 2100, for mean TCRE, in optimizing scenarios without overshoot.
This threshold is lower in case of slower GDP growth and early start
to abatement with smaller $N$. It is also lower in case of smaller
risk-free interest rates. The threshold rises by an increasing amount
for each year of delay in starting abatement, because the optimal
abatement rate grows with time. }

\end{figure}

\section{Conclusions and discussion}

Global warming from CO\textsubscript{2} is roughly proportional to
cumulative emissions, and more generally independent of emissions
pathway (\citet{Herrington2014,Seshadri2017a}), and such \textquotedblleft path-independence\textquotedblright{}
gives rise to cumulative emissions accounting and the concept of carbon
budgets (\citet{Stocker2013,Matthews2018}), presenting a simplification
robust to climate modeling uncertainty. It also poses new questions
about economic analyses of climate change abatement, through reducing
CO\textsubscript{2} emissions, under an exogenous constraint on cumulative
emissions. 

Unlike a Pigouvian tax, one that internalizes the estimated future
damage costs of emitting activities (\citet{Pearce2003,Pindyck2019}),
by approximating the social cost of carbon (SCC) (Appendix 7), the
present approach takes the carbon tax value to be instrumental in
achieving stringent emissions goals (\citet{Rogelj2011,Rogelj2015,Millar2017}).\footnote{Thus, while using the notation $P$ to denote a carbon tax, it is
no longer Pigouvian.} These, as well as the Paris Agreement (\citet{UNFCCC2015}), the
context for discussion of global warming mitigation, do not make direct
appeal to estimates of the SCC (\citet{Meinshausen2009,Hoegh-Guldberg2018}),
which is notoriously difficult to estimate, although it is recognized
that the SCC is quite large and has been underestimated in the past
(see for e.g. \citet{Hope2006,Nordhaus2017,Ricke2018,Pindyck2019}).
Moreover, carbon taxes in the present approach are more straightforward
to calculate, as the discussion surrounding Eqs. (\ref{eq:p19-1})
and (\ref{eq:p66}) shows; the SCC requires computing discounted future
effects of a CO\textsubscript{2} pulse and variation with time (Appendix
7), whereas in the present approach the challenge of integrating climate
models is averted by the finding that global warming at time $t$
only depends on cumulative emissions until then. However, a price
must be paid: the carbon tax can be higher or lower than the SCC,
since the present approach doesn't restrain its economic efficiency.
For example, imagine a rapid increase in temperature or the damage
function that would portend rapid increase in the SCC around future
time $t^{*}$: the former approach would have carbon tax increasing
rapidly around $t^{*}$, whereas that based on a cumulative emissions
constraint has it growing roughly at the interest-rate. 

Such an approach requires simple models of abatement, and we have
developed a model based on an increasing marginal abatement cost curve
that is shifted down or scaled in the presence of endogenous learning,
through either an additive or multiplicative term or both. The model
is similar to that of \citet{Emmerling2019}, with explicit treatment
of the effects of economic growth and furthermore endogenous learning
on the abatement curve. As the economy grows, business as usual emissions
grow correspondingly, and this leads to expansion of abatement options
and horizontal stretching of the cost curve. The present work has
many points of commonality with previous studies examining economics
under a cumulative abatement constraint (\citet{Dietz2019,Emmerling2019,Ploeg2019}),
but there are new features in the treatment of endogenous learning
in addition to climate damages, and analysis for time-varying rates,
in addition to applications to problems relevant to abatement. The
main findings are summarized below:

\subsection*{Optimal growth rates of abatement and the carbon tax}
\begin{itemize}
\item When learning and damages are not counted, the abatement rate relative
to the maximum abatement rate, which we denoted as $\sigma\left(t\right)$,
has growth rate given by the ratio of the risk-free interest rate
and the exponent of the abatement cost curve. A higher exponent, i.e.
sharply rising, abatement cost curve makes the abatement rate grow
more slowly, limiting future rise in abatement costs. This is consistent
with previous studies, e.g. \citet{Emmerling2019}.
\item However, the carbon tax growth rate equation is unaffected by the
shape of the marginal abatement cost curve, following Hotelling's
rule, similarly to previous studies (\citet{Dietz2019,Emmerling2019}).
A sharply rising abatement cost curve gives rise to a higher initial
abatement rate for the same carbon tax, with the abatement rate growing
more slowly with time. 
\end{itemize}

\subsection*{Effects of climate damage costs and endogenous learning}
\begin{itemize}
\item If cost-minimization accounts for endogenous learning, the carbon
price grows at a smaller rate as future marginal costs are reduced. 
\item Damage costs affect the optimal abatement pathway, especially where
the goal specifies less stringent levels of cumulative abatement,
for which global warming and resulting damage costs are higher. Limiting
damage costs, in addition to abatement costs, requires a higher carbon
price early on and therefore the subsequent growth rate of the carbon
tax can be smaller. 
\item The effect of endogenous learning on the initial carbon tax depends
on whether it enters as an additive of multiplicative model, i.e.
whether it shifts the marginal abatement cost curve down or reduces
it by a constant factor. Additive learning models lead to higher initial
abatement rates, in turn requiring a higher initial carbon tax. Multiplicative
learning models lead to slow abatement initially, involving lower
starting carbon taxes. 
\item Additive models of learning influence the optimal abatement rate only
if the interest rate is non-zero, owing to increasing marginal costs
(Appendix 5). However they reduce the growth rate of the carbon tax
regardless, because benefits of cost reductions do not depend on the
interest rate. 
\item Empirical literature on learning curves generally points to a multiplicative
model, implying lower initial carbon taxes (compared to the no-learning
case) that grow at a smaller rate. Competition between multiplicative
learning and damage effects on the abatement rate is resolved according
to the stringency of the abatement scenario: higher dependence on
learning in high-abatement scenarios, and on damages in low-abatement
scenarios. 
\end{itemize}

\subsection*{Initial carbon tax and cost of delaying abatement}
\begin{itemize}
\item The initial carbon tax rises with the stringency of abatement and
the BAU emissions, and is larger for lower interest rates as these
lead to higher abatement rates early on. The importance of a low discount
rate for the stringency of early abatement was emphasized previously
by \citet{Emmerling2019}. 
\item Delaying the start of abatement is costly as we have assumed that
the marginal abatement cost curve is increasing with the abatement
rate. Each year of delay leads to an increase in the total cost of
abatement in present value terms at $t=0$. A positive interest rate
does not imply that abatement can be costlessly postponed in a cumulative
emissions framework, because this has been factored into the optimal
abatement pathways. As a result, the need for more rapid abatement
drives the cost of a delay. 
\item The growth rate of the present value cost is larger for lower interest
rate and GDP growth rates, and increases rapidly for each year of
delay. Lower interest rates induce higher abatement rates early on
and therefore lead to higher relative costs of delaying abatement.
Slower GDP growth leads to a smaller growth in future abatement, so
the relative costs of delaying abatement are high. 
\item The relative increase in the cost of delay by one year does not depend
on the abatement level, but increases rapidly with proximity to the
end of the time-horizon. 
\item Each year of delay in abatement, postponing a global carbon tax, requires
a corresponding increase in the initial value of the tax if abatement
begins the following year. The carbon tax growth rate for each year
of delay has two components: a Hotelling rate, and a contribution
identical to the growth rate in present value expenditures, because
the marginal cost of abatement is proportional to the average cost
when we assume that the abatement curve has zero intercept. 
\end{itemize}

\subsection*{Economics of avoiding temperature overshoot}
\begin{itemize}
\item Avoiding temperature overshoot is important for preventing climate
change commitments that are irreversible, such as loss of ice-sheets
and biodiversity. Owing to proportionality between global warming
and cumulative emissions, overshoot in CO\textsubscript{2}'s contribution
to global warming generally corresponds to net negative emissions.
Therefore, we estimated the global warming goal attainable without
overshoot, $\triangle T_{fut}^{*}$, via optimizing pathways having
net positive emissions throughout and reaching zero emissions only
at $t=T$. 
\item This constraint of no net-negative emissions limits the threshold
(i.e. minimum) $\triangle T_{fut}^{*}$. This threshold rises with
GDP growth and a delayed start to abatement, because optimal pathways
have the annual abatement rate growing at rates governed by the interest
rate and economic growth. Hence they can abate only a fraction of
future BAU emissions, and high cumulative emissions goals over short
time-horizons cannot be met without net negative emissions. 
\item The threshold is reduced for smaller risk-free interest rates, and
high growth of the abatement cost, because these give rise to higher
abatement rates early on. For long time-horizons assuming an early
start to abatement, roughly the fraction $\theta r/\left(\frac{i}{c_{2}}+\theta r\right)$
of BAU emissions is abated. Only in the limit of zero interest rates
can all of BAU emissions be abated by stringent optimizing trajectories. 
\item Each year of delay in starting abatement makes the threshold larger
by an increasing amount, with marginal effects measured as $\partial\triangle T_{fut}^{*}/\partial N$,
rising as long as the optimal pathway has annual abatement rate growing. 
\item Overshoot increases by a small amount when effects of (multiplicative)
learning are included, and decreases by a small amount when economic
costs of climate damages are counted in the optimization. This effect
might increase if the damage fraction considers delayed effects, but
this has not been modeled. 
\end{itemize}
Creating yet more optimization-based models of abatement without constraint
from conservation principles or economic data is risk prone, especially
in studying long-term futures. At the same time, cumulative emissions
budgets pose an important constraint on global warming abatement,
whose implications for economic optimization approaches are an important
research area, leading to a growing body of work (\citet{Dietz2019,Emmerling2019}).
This paper has sought to develop a simplified model to study the satisfaction
of a cumulative emissions constraint at the lowest cost and, although
the resulting thought experiment is quite rudimentary, there are some
points of agreement with prior studies as noted above. There are also
several simplifications, which future work could remedy, especially:
making detailed estimates from data on costs, learning and damages,
with effects of the integral of cumulative abatement also included
to account for delayed effects such as those of sea-level rise; allowing
for epistemic uncertainty in future economic growth and interest rates;
and invoking further intertemporal choice criteria in studying decarbonization
pathways. 

\pagebreak{}

\section*{Appendix 1: Emissions intensity under business as usual}

Emissions intensity under business as usual at $t+\triangle t$ can
be written as $\frac{\dot{E}_{BAU}+\triangle\dot{E}_{BAU}}{G+\triangle G}=\frac{1+\theta\frac{\triangle G}{G}}{1+\frac{\triangle G}{G}}\mu$,
and expanding the denominator as its Taylor series

\begin{equation}
\frac{\dot{E}_{BAU}+\triangle\dot{E}_{BAU}}{G+\triangle G}=\left\{ 1+\left(\theta-1\right)\left(\left(\frac{\triangle G}{G}\right)-\left(\frac{\triangle G}{G}\right)^{2}+\ldots\right)\right\} \mu\label{eq:p34}
\end{equation}
The change is $\triangle\mu=\mu\left(t+\Delta t\right)-\mu\left(t\right)$,
and considering the natural increment of $\triangle t=1$ year so
that $\Delta G/G\equiv r$, the growth rate of global GDP, we obtain
approximately 
\begin{equation}
\frac{d\mu}{dt}\cong\frac{\triangle\mu}{\triangle t}=-\left(1-\theta\right)\left(r-r^{2}+r^{3}-\ldots\right)\mu\label{eq:p35}
\end{equation}

We might simplify the expression using the fact $r-r^{2}+r^{3}-\ldots$
is an infinite geometric series equaling $r/\left(1+r\right)$, so
the evolution of emissions intensity under BAU is $\frac{d\mu}{dt}\cong-\left(1-\theta\right)\frac{r}{1+r}\mu$.

\section*{Appendix 2: Derivation of Euler-Lagrange equation}

For fixed $T$, in order to minimize the integral $\int_{0}^{T}g\left(t,M\left(t\right),\dot{M}\left(t\right)\right)dt$
subject to constraint $\int_{0}^{T}m\left(t\right)dt=M_{tot}$, we
first define $I\left(M\left(t\right),\dot{M}\left(t\right)\right)=\int_{0}^{T}g\left(t\right)dt+\lambda\left(\int_{0}^{T}\dot{M}\left(t\right)dt-M_{tot}\right)$.
Stationary requires the first-variation $\delta I$ in the integral
due to small changes $\delta M$ and $\delta\dot{M}$ to vanish. The
variation $\delta I=I\left(M+\delta M,\dot{M}+\delta\dot{M}\right)-I\left(M,\dot{M}\right)$
is 
\begin{equation}
\delta I=\int_{0}^{T}\left\{ \frac{\partial g}{\partial M}\delta M+\frac{\partial g}{\partial\dot{M}}\delta\dot{M}+\lambda\left(\frac{\partial\dot{M}}{\partial M}\delta M+\frac{\partial\dot{M}}{\partial\dot{M}}\delta\dot{M}\right)\right\} dt=0\label{eq:p36}
\end{equation}
and, integrating by parts, $\int_{0}^{T}\frac{\partial g}{\partial\dot{M}}\delta\dot{M}dt=-\int_{0}^{T}\frac{d}{dt}\left(\frac{\partial g}{\partial\dot{M}}\right)\delta Mdt$
after applying the condition that $\delta M\left(0\right)=\delta M\left(T\right)=0$
because the variations must preserve boundary conditions of the problem.
There is a corresponding equation involving the integral constraint
on cumulative negative emissions. Combining these equations yields
\begin{equation}
\delta I=\int_{0}^{T}\left\{ \frac{\partial g}{\partial M}-\frac{d}{dt}\left(\frac{\partial g}{\partial\dot{M}}\right)+\lambda\left(\frac{\partial\dot{M}}{\partial M}-\frac{d}{dt}\left(\frac{\partial\dot{M}}{\partial\dot{M}}\right)\right)\right\} \delta Mdt=0\label{eq:p37}
\end{equation}
 for arbitrary changes $\delta M$, yielding the Euler-Lagrange Eq.
(\ref{eq:p12}).

\section*{Appendix 3: Terms in Euler-Lagrange equation }

From Eqs. (\ref{eq:p11}) and (\ref{eq:p13})
\begin{multline}
\frac{\partial g}{\partial M}=-e^{-I\left(t\right)}f'\left(M\left(t\right)\right)h\left(M\left(t\right)\right)\dot{M}\left(t\right)+e^{-I\left(t\right)}\left\{ c_{0}+c_{1}\left(\frac{\dot{M}\left(t\right)}{\dot{M}_{max}\left(t\right)}\right)^{c_{2}}-f\left(M\left(t\right)\right)\right\} h'\left(M\left(t\right)\right)\dot{M}\left(t\right)\\
+e^{-I\left(t\right)}d'\left(M\left(t\right)\right)G\left(t\right)\label{eq:p38}
\end{multline}
whereas 
\begin{multline}
\frac{\partial g}{\partial\dot{M}}=e^{-I\left(t\right)}c_{1}c_{2}\left(\frac{\dot{M}\left(t\right)}{\dot{M}_{max}}\right)^{c_{2}}h\left(M\left(t\right)\right)+e^{-I\left(t\right)}\left\{ c_{0}+c_{1}\left(\frac{\dot{M}\left(t\right)}{\dot{M}_{max}\left(t\right)}\right)^{c_{2}}-f\left(M\left(t\right)\right)\right\} h\left(M\left(t\right)\right)\label{eq:p39}
\end{multline}
Here we have assumed that $\dot{M}_{max}\left(t\right)$ is a known
function of time, which can be related to BAU emissions, and hence
independent of $M\left(t\right)$ and $\dot{M}\left(t\right)$. The
above equation simplifies to
\begin{multline}
\frac{\partial g}{\partial\dot{M}}=e^{-I\left(t\right)}\left\{ c_{0}+c_{1}\left(c_{2}+1\right)\left(\frac{\dot{M}\left(t\right)}{\dot{M}_{max}\left(t\right)}\right)^{c_{2}}-f\left(M\left(t\right)\right)\right\} h\left(M\left(t\right)\right)\label{eq:p40}
\end{multline}
so that 
\begin{multline}
\frac{d}{dt}\left(\frac{\partial g}{\partial\dot{M}}\right)=-i\left(t\right)e^{-I\left(t\right)}\left\{ c_{0}+c_{1}\left(c_{2}+1\right)\left(\frac{\dot{M}\left(t\right)}{\dot{M}_{max}\left(t\right)}\right)^{c_{2}}-f\left(M\left(t\right)\right)\right\} h\left(M\left(t\right)\right)\\
+e^{-I\left(t\right)}\left\{ \frac{c_{1}c_{2}\left(c_{2}+1\right)}{\dot{M}_{max}^{c_{2}}}\dot{M}\left(t\right)^{c_{2}-1}\ddot{M}\left(t\right)-f'\left(M\left(t\right)\right)\dot{M}\left(t\right)\right\} h\left(M\left(t\right)\right)\\
+e^{-I\left(t\right)}\left\{ c_{0}+c_{1}\left(c_{2}+1\right)\left(\frac{\dot{M}\left(t\right)}{\dot{M}_{max}\left(t\right)}\right)^{c_{2}}-f\left(M\left(t\right)\right)\right\} h'\left(M\left(t\right)\right)\dot{M}\left(t\right)\\
-e^{-I\left(t\right)}c_{1}c_{2}\left(c_{2}+1\right)\left(\frac{\dot{M}\left(t\right)}{\dot{M}_{max}\left(t\right)}\right)^{c_{2}}\frac{1}{\dot{M}_{max}\left(t\right)}\left\{ \frac{d}{dt}\dot{M}_{max}\left(t\right)\right\} h\left(M\left(t\right)\right)\label{eq:p41}
\end{multline}
 Using $\frac{\partial\dot{M}}{\partial M}=0$ and $\frac{\partial\dot{M}}{\partial\dot{M}}=1$,
so that $\frac{d}{dt}\left(\frac{\partial\dot{M}}{\partial\dot{M}}\right)=0$
we finally obtain
\begin{multline}
d'\left(M\left(t\right)\right)G\left(t\right)=-i\left(t\right)\left\{ c_{0}+c_{1}\left(c_{2}+1\right)\left(\frac{\dot{M}\left(t\right)}{\dot{M}_{max}}\right)^{c_{2}}-f\left(M\left(t\right)\right)\right\} h\left(M\left(t\right)\right)\\
+\frac{c_{1}c_{2}\left(c_{2}+1\right)}{\dot{M}_{max}^{c_{2}}}\dot{M}\left(t\right)^{c_{2}-1}\ddot{M}\left(t\right)h\left(M\left(t\right)\right)+\frac{c_{1}c_{2}}{\dot{M}_{max}^{c_{2}}}\dot{M}\left(t\right)^{c_{2}+1}h'\left(M\left(t\right)\right)\\
-c_{1}c_{2}\left(c_{2}+1\right)\left(\frac{\dot{M}\left(t\right)}{\dot{M}_{max}\left(t\right)}\right)^{c_{2}}\frac{1}{\dot{M}_{max}\left(t\right)}\left\{ \frac{d}{dt}\dot{M}_{max}\left(t\right)\right\} h\left(M\left(t\right)\right)\label{eq:p44}
\end{multline}
 and taking the 2\textsuperscript{nd} derivative term to one side
and dividing by $h\left(t\right)>0$ yields Eq. (\ref{eq:p14}). 

\section*{Appendix 4: Effects on marginal cost curve of the learning models}

Let us consider the influence of the two models of endogenous learning
on the marginal cost curve. Average cost $\gamma\left(\dot{M},M\right)$
is related to marginal cost $\beta\left(M,\dot{M}\right)$ as $\gamma\left(M,\dot{M}\right)=\frac{1}{\dot{M}}\int_{0}^{\dot{M}}\beta\left(M,\dot{s}\right)d\dot{s}$
where $\dot{s}$ is integrated from $0$ to $\dot{M}$ . Substituting
for the form of the average cost in Eq. (\ref{eq:p3}), marginal cost
is related as
\begin{equation}
\int_{0}^{\dot{M}}\beta\left(M,\dot{s}\right)d\dot{s}=\left(c_{0}+c_{1}\left(\frac{\dot{M}\left(t\right)}{\dot{M}_{max}}\right)^{c_{2}}-f\left(M\right)\right)h\left(M\right)\dot{M}\label{eq:p45}
\end{equation}
and differentiating with respect to $\dot{M}$ 
\begin{equation}
\beta\left(M,\dot{M}\right)=\left\{ c_{0}+\frac{\left(c_{2}+1\right)c_{1}}{\dot{M}_{max}^{c_{2}}}\dot{M}^{c_{2}}-f\left(M\right)\right\} h\left(M\right)\label{eq:p46}
\end{equation}
which makes it clear that $f\left(M\right)$ is an additive contribution
to the marginal cost curve whereas $h\left(M\right)$ multiplies it
throughout. 

\section*{Appendix 5: Effect of additive endogenous learning with zero interest
rate}

The need for nonzero interest rate $i\left(t\right)$, for a non-zero
effect of the additive learning model on the optimal abatement pathway,
arises even if we assume that there is no GDP growth, wherein $r_{BAU}\left(t\right)=0$
and $\dot{M}_{max}$ is constant. Therefore we make these restrictions
in our discussion of origins of this behavior. Consider abatement
rate $\dot{M}\left(t\right)=\frac{M_{tot}}{T}+\epsilon x\left(t\right)$,
with $0<\epsilon\ll\frac{M_{tot}}{T}$,\footnote{We assume $\max\left|x\left(t\right)\right|$ is of the order $1$,
so $\epsilon x\left(t\right)$ is small in magnitude compared to $\frac{M_{tot}}{T}$.} so that $\epsilon x\left(t\right)$ is a small perturbation to a
trajectory with constant abatement rate. Cumulative abatement rate
in this case is $M\left(t\right)=\int_{0}^{t}\dot{M}\left(s\right)ds=M_{tot}\frac{t}{T}+\epsilon X\left(t\right)$,
where $X\left(t\right)=\int_{0}^{t}x\left(s\right)ds$. Boundary conditions
on $M\left(t\right)$ require $X\left(0\right)=X\left(T\right)=0$.
If only the additive model is $f\left(M\left(t\right)\right)$ is
active, with $h\left(M\left(t\right)\right)$ being fixed at $1$,
the abatement cost in year $t$ is, from Eq. (\ref{eq:p9}), after
substituting the models of $M\left(t\right)$ and $\dot{M}\left(t\right)$
above
\begin{multline}
A\left(t\right)=c_{0}\left\{ \frac{M_{tot}}{T}+\epsilon x\left(t\right)\right\} +\frac{c_{1}}{\dot{M}_{max}}\left(\frac{M_{tot}}{T}+\epsilon x\left(t\right)\right)^{c_{2}+1}\\
-\left\{ \frac{M_{tot}}{T}+\epsilon x\left(t\right)\right\} f\left(M_{tot}\frac{t}{T}+\epsilon X\left(t\right)\right)\label{eq:p47}
\end{multline}
We have assumed that $\dot{M}_{max}$ is constant, as in a steady-state
GDP trajectory, as this simplifies our discussion. As noted above,
the effect of the interest rate on additive learning does not depend
on any growth assumptions. We furthermore approximate the function
$f$ above by its Taylor series truncated to $1$\textsuperscript{st}-order
and following some algebra, and counting only abatement costs, the
discounted expenditure in case $i=0$ simplifies to
\begin{equation}
C_{total}=\int_{0}^{T}A\left(t\right)dt=C_{0}M_{tot}-\frac{M_{tot}}{T}\int_{0}^{T}f\left(M_{tot}\frac{t}{T}\right)dt+\frac{c_{1}}{\dot{M}_{max}}\int_{0}^{T}\dot{M}\left(t\right)^{c_{2}+1}dt\label{eq:p48}
\end{equation}
as described below. Those uninterested in the details might skip the
aside below to the discussion following Eq. (\ref{eq:p53}), where
the thread of the present discussion is resumed. 

\{Aside: In the following we derive Eq. (\ref{eq:p48}). For abatement
expenditure in year $t$ in Eq. (\ref{eq:p47}), upon substituting
the Taylor series expansion to $1$\textsuperscript{st}order in small
parameter $\epsilon$
\begin{equation}
f\left(M_{tot}\frac{t}{T}+\epsilon X\left(t\right)\right)\cong f\left(M_{tot}\frac{t}{T}\right)+\epsilon X\left(t\right)f'\left(M_{tot}\frac{t}{T}\right)\label{eq:p49}
\end{equation}
we obtain
\begin{multline}
A\left(t\right)=\frac{M_{tot}}{T}\left\{ c_{0}-f\left(M_{tot}\frac{t}{T}\right)\right\} +\epsilon\frac{d}{dt}\left\{ c_{0}X\left(t\right)-f\left(M_{tot}\frac{t}{T}\right)X\left(t\right)\right\} \\
+\frac{c_{1}}{\dot{M}_{max}}\left(\frac{M_{tot}}{T}+\epsilon x\left(t\right)\right)^{c_{2}+1}\label{eq:p50}
\end{multline}
where, for consistency, we have retained only $1$\textsuperscript{st}
order terms in $\epsilon$ for the series-derived contributions. Total
expenditure becomes, for $\delta=0$
\begin{multline}
C_{total}=\int_{0}^{T}\frac{M_{tot}}{T}\left\{ c_{0}-f\left(M_{tot}\frac{t}{T}\right)\right\} dt+\\
\epsilon\int_{0}^{T}\frac{d}{dt}\left\{ c_{0}X\left(t\right)-f\left(M_{tot}\frac{t}{T}\right)X\left(t\right)\right\} dt+\frac{c_{1}}{\dot{M}_{max}}\int_{0}^{T}\left(\frac{M_{tot}}{T}+\epsilon x\left(t\right)\right)^{c_{2}+1}dt\label{eq:p51}
\end{multline}
and, using
\begin{multline}
\int_{0}^{T}\frac{d}{dt}\left\{ c_{0}X\left(t\right)-f\left(M_{tot}\frac{t}{T}\right)X\left(t\right)\right\} dt=\\
c_{0}\left(X(T)-X\left(0\right)\right)+f\left(T_{tot}\right)X\left(T\right)-f\left(0\right)X\left(0\right)=0\label{eq:p52}
\end{multline}
with the last equality following from boundary conditions on $X\left(t\right)$,
we obtain
\begin{equation}
C_{total}=\int_{0}^{T}\frac{M_{tot}}{T}\left\{ c_{0}-f\left(M_{tot}\frac{t}{T}\right)\right\} dt+\frac{c_{1}}{\dot{M}_{max}}\int_{0}^{T}\left(\frac{M_{tot}}{T}+\epsilon x\left(t\right)\right)^{c_{2}+1}dt\label{eq:p53}
\end{equation}
which simplifies to Eq. (\ref{eq:p48}).\}

It has been shown in the Aside that the $1$\textsuperscript{st}-order
effect of an arbitrary additive learning function on a constant emissions
trajectory is zero. Moreover, the effect of non-constant abatement
rate enters only through the last term in Eq. (\ref{eq:p48}), which
we expect to be minimized by the constant abatement rate pathway.
This is shown in the following. Since $c_{2}>0$, $p\left(\dot{M}\right)\equiv\dot{M}^{c_{2}+1}$
is a convex function of $\dot{M}$. Recall that emissions rate $\dot{M}\left(t\right)=\frac{M_{tot}}{T}+\epsilon x\left(t\right)$,
with its mean value $\mathbf{E}\dot{M}=\frac{1}{T}\int_{0}^{T}\dot{M}\left(t\right)dt=\frac{M_{tot}}{T}$
and, furthermore
\begin{equation}
\mathbf{E}p\left(\dot{M}\right)=\frac{1}{T}\int_{0}^{T}p\left(\dot{M}\left(t\right)\right)dt=\frac{1}{T}\int_{0}^{T}\dot{M}\left(t\right)^{c_{2}+1}dt\label{eq:p54}
\end{equation}
From Jensen's inequality, for a convex function $p\left(\dot{M}\right)$,
$\mathbf{E}p\left(\dot{M}\right)\geq p\left(\mathbf{E}\dot{M}\right)$.
But $p\left(\mathbf{E}\dot{M}\right)=\left(\frac{M_{tot}}{T}\right)^{c_{2}+1}$.
Hence
\begin{equation}
\frac{1}{T}\int_{0}^{T}\dot{M}\left(t\right)^{c_{2}+1}dt\geq\left(\frac{M_{tot}}{T}\right)^{c_{2}+1}\label{eq:p55}
\end{equation}
or equivalently
\begin{equation}
\frac{c_{1}}{\dot{M}_{max}}\int_{0}^{T}\left(\frac{M_{tot}}{T}+\epsilon x\left(t\right)\right)^{c_{2}+1}dt\geq\frac{c_{1}}{\dot{M}_{max}}\int_{-T}^{0}\left(\frac{M_{tot}}{T}\right)^{c_{2}+1}dt\label{eq:p56}
\end{equation}
with equality occurring at $\epsilon=0$. Minimum entails $\epsilon=0$,
so that the additive endogenous learning model does not influence
the optimal path if the risk-free interest rate is zero. This result
appeals to convexity of function $\dot{N}\left(t\right)^{c_{2}+1}$,
but this does not limit its generality because the function is surely
convex if $c_{2}>0$, where marginal costs are increasing.

\section*{Appendix 6: Present value of abatement cost without endogenous learning}

Assuming the absence of endogenous learning, the present value of
the abatement cost is 
\begin{equation}
C_{total}=c_{1}\int_{N}^{T}e^{-I\left(t\right)}\sigma\left(t\right)^{c_{2}+1}\dot{M}_{max}\left(t\right)dt\label{eq:p57}
\end{equation}
assuming $c_{0}=0$, and using Eqs. (\ref{eq:p2}) and (\ref{eq:p15}).
For this situation, from Eq. (\ref{eq:p16}) $\sigma\left(t\right)=\sigma\left(N\right)e^{\int_{N}^{t}\frac{i\left(s\right)}{c_{2}}ds}=\sigma\left(N\right)e^{-I\left(N\right)/c_{2}}e^{I\left(t\right)/c_{2}}$,
where we recall that $I\left(t\right)=\int_{0}^{t}i\left(s\right)ds$.
Similarly, $\dot{M}_{max}\left(t\right)=\dot{M}_{max}\left(N\right)e^{\int_{N}^{t}\theta r\left(s\right)ds}$
can also be written as $\dot{M}_{max}\left(t\right)=\dot{M}_{max}\left(N\right)e^{-\theta R\left(N\right)}e^{\theta R\left(t\right)}$,
whereupon 
\begin{equation}
C_{total}=c_{1}\sigma\left(N\right)^{c_{2}+1}\dot{M}_{max}\left(N\right)e^{-I\left(N\right)\left(1+\frac{1}{c_{2}}\right)}e^{-\theta R\left(N\right)}\int_{N}^{T}e^{I\left(t\right)/c_{2}+\theta R\left(t\right)}dt\label{eq:p58}
\end{equation}
or equivalently
\begin{equation}
C_{total}=C\left(N\right)e^{-I\left(N\right)\left(1+\frac{1}{c_{2}}\right)}e^{-\theta R\left(N\right)}\int_{N}^{T}e^{I\left(t\right)/c_{2}+\theta R\left(t\right)}dt\label{eq:p59}
\end{equation}
where 
\begin{equation}
C\left(N\right)=c_{1}\sigma\left(N\right)^{c_{2}+1}\dot{M}_{max}\left(N\right)\label{eq:p60}
\end{equation}
is the abatement cost during the initial year $t=N$. Its value is
estimated from the cumulative abatement constraint. The evolution
of the abatement rate is $\dot{M}\left(t\right)=\dot{M}\left(N\right)e^{\int_{N}^{t}\left\{ \frac{i\left(s\right)}{c_{2}}+\theta r\left(s\right)\right\} ds}$,
so the cumulative abatement becomes 
\begin{equation}
M_{tot}=\int_{N}^{T}\dot{M}\left(t\right)dt=\dot{M}\left(N\right)e^{-I\left(N\right)/c_{2}}e^{-\theta R\left(N\right)}\int_{N}^{T}e^{I\left(t\right)/c_{2}+\theta R\left(t\right)}dt\label{eq:p61}
\end{equation}
from which $\sigma\left(N\right)=\dot{M}\left(N\right)/\dot{M}_{max}\left(N\right)$
reduces to
\begin{equation}
\sigma\left(N\right)=\frac{M_{tot}}{\dot{M}_{max}\left(0\right)}\frac{e^{I\left(N\right)/c_{2}}}{\int_{N}^{T}e^{I\left(t\right)/c_{2}+\theta R\left(t\right)}dt}\label{eq:p62}
\end{equation}
where we have used $\dot{M}_{max}\left(N\right)=\dot{M}_{max}\left(0\right)e^{\theta R\left(N\right)}$.
Substituting into Eq. (\ref{eq:p60})
\begin{equation}
C\left(N\right)=c_{1}\left(\frac{M_{tot}}{\dot{M}_{max}\left(0\right)}\right)^{c_{2}+1}\dot{M}_{max}\left(0\right)\frac{e^{I\left(N\right)\left(1+\frac{1}{c_{2}}\right)}e^{\theta R\left(N\right)}}{\left\{ \int_{N}^{T}e^{I\left(t\right)/c_{2}+\theta R\left(t\right)}dt\right\} ^{c_{2}+1}}\label{eq:p63}
\end{equation}
so that finally, from Eq. (\ref{eq:p59})
\begin{equation}
C_{total}=c_{1}\left(\frac{M_{tot}}{\dot{M}_{max}\left(0\right)}\right)^{c_{2}+1}\dot{M}_{max}\left(0\right)\frac{1}{\left\{ \int_{N}^{T}e^{I\left(t\right)/c_{2}+\theta R\left(t\right)}dt\right\} ^{c_{2}}}\label{eq:p64}
\end{equation}

\section*{Appendix 7: Social cost of carbon}

The social cost of carbon (SCC) is usually defined as the discounted
damage costs from a pulse emission of CO\textsubscript{2}, conventionally
one ton
\begin{equation}
SCC\left(t\right)=\int_{t}^{\infty}e^{-\tilde{I}\left(t\right)}d'\left(T\left(s\right)\right)G\left(s\right)\triangle\widetilde{T}\left(s\right)ds\label{eq:p65}
\end{equation}
where the SCC is discounted at a rate $\tilde{i}\left(t\right)$ that
in general is distinct from the risk-free interest rate. The SCC is
calculated assuming that a unit pulse of CO\textsubscript{2} is emitted
at time $t$, and depends on the subsequent global warming contribution
from that pulse $\triangle\widetilde{T}\left(s\right)$, $s\geq t$,
with $\triangle\widetilde{T}\left(s=t\right)=0$; and on the sensitivity
of global climate damages to temperature changes $d'\left(T\left(s\right)\right)G\left(s\right)$.
For example, for a power-law model of damage fraction the social cost
of carbon is
\begin{equation}
SCC\left(t\right)=\frac{d_{1}}{TCRE}\int_{t}^{\infty}e^{-\tilde{I}\left(t\right)}\frac{D\left(s\right)}{E\left(s\right)}\triangle\widetilde{T}\left(s\right)ds\label{eq:p66}
\end{equation}
where, we recall, TCRE is the constant ratio of global warming to
cumulative emissions, $d_{1}>1$ is the exponent of the damage function,
and $D\left(s\right)/E\left(s\right)$ is the damage costs per cumulative
emissions. Since damages are a convex function of global warming,
this ratio is increasing as long as global temperature is rising.
This is the main factor behind increasing SCC with time. The contribution
$\triangle\widetilde{T}\left(s\right)$ depends on the competition
between the diminishing radiative forcing from a unit pulse of CO\textsubscript{2},
owing the logarithmic radiative forcing relation with excess concentrations,
and the decreasing ability of the oceans to take up excess heat and
CO\textsubscript{2}, as global warming proceeds. Generally $\triangle\widetilde{T}\left(s\right)$
is not merely $\triangle\widetilde{T}\left(s-t\right)$ , but actually
$\triangle\widetilde{T}\left(s-t,\triangle T\left(s\right),CO_{2}\left(s\right)\right),$i.e.
depending not only on the time elapsed since the emission of the pulse,
but also global warming $\triangle T\left(s\right)$ in the absence
of the pulse which affects its fate in the atmosphere, as well as
excess carbon-dioxide in the atmosphere $CO_{2}\left(s\right)$ in
the absence of the pulse which affects the radiative forcing contribution
from the pulse (\citet{Seshadri2017}). Furthermore, the ratio $D\left(s\right)/E\left(s\right)$
appearing in the integral, for times $s\geq t$, requires not only
present values of global warming and cumulative emissions, but also
assumptions about future evolution of cumulative emissions so that
damages can be calculated for all times $s\geq t$. Clearly, the information
needs for estimating the SCC are onerous and corresponding uncertainties
are large.\pagebreak{}

\bibliographystyle{../agufull08}
\bibliography{../econ1}

\end{document}